\documentclass[preprint,12pt]{aastex}
\usepackage{graphicx}
\usepackage{natbib}
\usepackage{amsmath}
\shortauthors{INOUE, SAKAI, and TOMITA}
\shorttitle{Evidence of Quasi-linear Super-Structures in 
the CMB and Galaxy Distribution}

\newcommand{\f}{\frac}
\newcommand{\T}{\tilde}

\newcommand{\bb}{\bibitem}
\newcommand{\BF}{\begin{figure}\begin{center}}
\newcommand{\EF}{\end{center}\end{figure}}
\newcommand{\BE}{\begin{equation}}
\newcommand{\EE}{\end{equation}}
\newcommand{\BEA}{\begin{eqnarray}}
\newcommand{\EEA}{\end{eqnarray}}

\begin{document}
\title{Evidence of Quasi-linear Super-Structures in 
the Cosmic Microwave Background and Galaxy Distribution}
\author{Kaiki Taro Inoue\altaffilmark{1}, Nobuyuki Sakai\altaffilmark{2} 
and Kenji Tomita\altaffilmark{3}}
\altaffiltext{1}{Department of Science and Engineering, 
Kinki University, Higashi-Osaka, 577-8502, Japan}
\altaffiltext{2}{Department of Education, Yamagata University, 
Yamagata 990-8560, Japan}
\altaffiltext{3}{Yukawa Institute for Theoretical Physics, 
Kyoto University, Kyoto 606-8502, Japan}


\begin{abstract}
Recent measurements of hot and cold spots 
on the cosmic microwave background (CMB) sky suggest a presence of 
super-structures on ($>100\,h^{-1}$Mpc) scales.  We develop a new
formalism to estimate the expected amplitude of temperature fluctuations 
due to the integrated Sachs-Wolfe (ISW) effect from prominent 
quasi-linear structures.
Applying the developed tools to the observed ISW signals from 
voids and clusters in catalogs of galaxies at redshifts $z<1$, 
we find that they indeed 
imply a presence of quasi-linear super-structures
with a comoving radius $100\sim 300\,h^{-1}$Mpc and a
density contrast $|\delta|\sim O(0.1)$.
We also find that the observed ISW signals are at odd with the concordant
$\Lambda$ cold dark matter (CDM) model that predicts Gaussian primordial perturbations
at $\gtrsim 3\, \sigma$ level. We confirm that the mean temperature
around the CMB cold spot in the 
southern Galactic hemisphere filtered by 
a compensating top-hat filter deviates from a mean value at 
$\sim 3\,\sigma$ level, implying that a quasi-linear 
supervoid or an underdensity region surrounded by a massive wall 
may reside at low redshifts $z<0.3$ and the actual 
angular size ($16^{\circ}-17^\circ$) may be larger than the apparent
size ($4^\circ-10^\circ$) discussed in literature.  
Possible solutions are briefly discussed.

\end{abstract}


\section{Introduction}
Although the $\Lambda$ cold dark matter (CDM) models have 
succeeded in explaining a number of observations, 
some problems remain unresolved.
For example, origins of a possible break of 
statistical isotropy in the large-angle
cosmic microwave background (CMB) anisotropy (Tegmark et al. 
2003; Eriksen et al. 2004; Vielva et al. 2004) and  
a possible discrepancy between observed 
and theoretically predicted  
galaxy-CMB cross-correlation (Rassat et al. 2007; Ho et al. 2008)
are still not understood well.  
These observational results imply that  
structures on scales larger than $\gtrsim 100$ Mpc (super-structures) 
in our local universe are more lumpy than expected (Afshordi et al. 2009).

As the origin of the large-angle CMB anomalies, many authors
have considered a possibility that the CMB is affected by 
local inhomogeneities (Moffat 2005; 
Tomita 2005a,b; Cooray \& Seto 2005; Rakic \& Schwartz 2007).  
Inoue \& Silk (2006, 2007) have shown that 
a particular configuration of compensated quasi-linear 
supervoids can explain most of the features of the anomalies. 
Subsequent theoretical analyses have shown that the CMB temperature
distribution for quasi-linear structures can be 
skewed toward low temperature due to
the second order integrated Sachs-Wolfe (ISW)(or Rees-Sciama) 
effect (Tomita \& Inoue 2008; Sakai \& Inoue 2008).  

In fact, Granett et al. (2008) found 
a significant ISW signal at the scale of 
$4^\circ-6^\circ$ at redshifts around $z \sim 0.5$ and a weak 
signal of negatively skewed temperature 
distribution for distinct voids and clusters at redshifts $0.4<z<0.75$.  
Moreover, 
Francis \& Peacock (2009)
have shown that the ISW effect due to local structures at redshift $z<0.3$ 
significantly affects the large-angle CMB anisotropies and that 
some of the CMB anomalies no longer persist after subtraction of the ISW 
contribution.  These observations 
of galaxy-CMB cross-correlation 
may suggest an existence of anomalously large perturbations 
or new physics on scales $>100$ Mpc.  

In order to evaluate the significance of the ISW signals 
for prominent structures, 
N-body simulations on cosmological scales seem to be
suitable(Cai et al. 2010) for this purpose. However, the 
computation time is relatively long and finding 
physical interpretation from a number of numerical results 
is sometimes difficult. In contrast, 
analytical methods are suitable for estimating the order of   
statistical significance in a relatively short time, and 
physical interpretations are often simpler. 

In this paper, we evaluate the statistical significance of 
the ISW signals for prominent super-structures based on 
an analytic method and try to construct simple models that 
are consistent with the data. 
In section 2, we develop a formalism for analytically 
calculating the ISW signal
due to prominent non-linear super-structures based on a spherically symmetric
homogeneous collapse model and we study the effect of non-linearity
and inhomogeneity of such structures.  In section 3, we apply the 
developed method to observed data and 
calculate the statistical significance of the discrepancy 
between the predicted and the observed ISW signals.  
In section 4, we discuss the origin of the observed discrepancy. 
In section 5, we summarize our results and discuss some unresolved issues.
In the following, unless noted, we assume a 
concordant $\Lambda$CDM cosmology with $(\Omega_{m,0}, \Omega_{\Lambda,0}, \Omega_{b,0}, h,\sigma_8,n)=(0.26,0.74,0.044,0.72,0.80,0.96)$, which agrees with the recent CMB
and large-scale structure data (S\'anchez et al. 2009).
\pagebreak
\section{Cross correlation for prominent quasi-linear structures}
\subsection{Thin-shell approximation}

For simplicity, in this section, we assume that 
super-structures are modelled by spherically symmetric 
homogeneous compensated voids/clusters with an 
infinitesimally thin-shell. The background spacetime
is assumed to be a flat FRW universe with
matter and a cosmological constant $\Lambda$.

Let $\kappa$ and $\xi$ be the curvature and the physical radius 
of a void/cluster in unit of the Hubble radius $H^{-1}$, and 
$\delta_H$ as the Hubble parameter contrast, 
$t$ denotes the cosmological time. We describe the 
angle between the normal vector of the shell
and the three dimensional momentum of the CMB photon that 
leaves the shell by $\psi$. 
We assume that the comoving radius $r_v$ of the void in the 
background coordinates satisfies $r_v \propto t^\beta$, where 
$\beta$ is a constant.

Up to order $O((r_v/H^{-1})^3)$
and $O(\kappa^2)$, the temperature anisotropy of the CMB photons that pass
through spherical homogeneous compensated 
voids in the flat FRW universe can be written as (Inoue \& Silk 2007), 
\BEA
\f{\it{\Delta T}}{T}&=&\f{1}{3 }\biggl[\xi^3 \cos{\psi}
\Bigl(-2 \delta_H^2-\delta_H^3+(3+4 \delta_m)\delta_H \Omega_m
\nonumber
\\
\nonumber
&+&\delta_m \Omega_m(-6 \beta/s +1)+
(2\delta_H^2+\delta_H^3+\delta_m \Omega_m
\\
&+&(3+ 2\delta_m)
\delta_H \Omega_m)\cos{2 \psi}\Bigr)\biggr]
,
\label{eq:dToverTthin}
\\
t&=&s H^{-1}, ~s=\f{2}{3\sqrt{1-\Omega_m}} 
 \ln 
\biggl[\f{\sqrt{1-\Omega_m}+1}{\sqrt{\Omega_m}}   \biggr],
\\
\nonumber
\EEA
where $\delta_m$ and $\Omega_m$
denote the matter density contrast of the void 
and the matter density parameter, respectively. The variables $\xi, \psi,
\delta_m,\delta_H$, and $\Omega_m$ 
are evaluated at the time the CMB photon leaves the shell. 
It should be noted that the formula (\ref{eq:dToverTthin})
is valid even if $|\delta_m|$ or $|\delta_H|$ is somewhat large
as long as the normalized curvature $\kappa$ is
sufficiently small. 
The formula (\ref{eq:dToverTthin}) can be also
applied to spherical compensating clusters with a density contrast 
$\delta_m>0$ corresponding to 
a homogeneous spherical cluster with an infinitesimally thin ``wall''. 
This approximation holds 
only in weakly non-linear regime since the amplitude of 
the density contrast corresponding to a negative mass 
cannot exceed 1. We examine this approximation 
in sections 2-4 in detail.

Because we are mainly
interested in linear $|\delta_m|\ll 1$ and quasi-linear
$|\delta_m|=O(0.1)$
regime, we expand $\delta_H$ in terms of $\delta_m$ up to second order 
as 
\BE
\delta_H=\Omega_m \delta_m (1+1/f(w))/2-\epsilon \delta_m^2, 
\label{deltaH}
\EE
where $w$ is an equation-of-state parameter, $\epsilon$ is 
a constant that describes the non-linear effect and 
\BE
f(w)=-\f{3}{5}(1+w)^{1/3}{}_2 F_1\biggl[ \f{5}{6},\f{1}{3},\f{11}{6},-w
\biggr ],
\EE
where  $_2 F_1$ is Gauss' hypergeometric function.
$\epsilon$ can be estimated from 
numerical integration of the Friedmann equation
inside the shell as we shall show later. 
In a similar manner, for the shell expansion, we assume 
the following relation for the wall peculiar velocity normalized by
the background Hubble expansion, 
\BE
\tilde{v}=s^{-1}\beta,~~
\beta =-\f{f(\Omega_m)}{6} \delta_m+\nu \delta_m^2,
\EE
where $\nu$ represents a constant 
that describes the non-linear effect (Inoue \& Silk 2007).
$f(\Omega_m)$ is written in terms of the scale factor $a$
and the growth factor $D$ as
\BE
f(\Omega_m)=\f{a}{D}\f{dD}{da}\sim \Omega_m^{0.6}.
\EE

In quasi-linear regime, simplification $\epsilon=\eta=0$ 
can be verified, which will be shown in section 2-3 and section 2-4. 
\subsection{Homogeneous collapse}
In order to describe the dynamics of local inhomogeneity,
we adopt a homogeneous collapse model which consists of
an inner FRW patch and a surrounding 
background flat FRW spacetime (Lahav et al. 1991). The size of the inner 
patch is assumed to be sufficiently smaller than the 
horizon $H^{-1}$ in the background spacetime.

We assume that both the regions
have only dust and a cosmological constant $\Lambda$. 
The time evolution of either the inner patch or 
the background spacetime
is described by the Friedmann equation,
\BE
\f{H^2}{H_0^2}=\f{\Omega_{m,0}}{a^3}+\Omega_{\Lambda,0}+\f{1-\Omega_{tot}}{a^2},
\label{eq:1}
\EE
where $a$ denotes the scale factor,
$\Omega_{m,0}, \Omega_{\Lambda,0}, \Omega_{tot}$ 
are the present energy density parameters of non-relativistic matter,
a cosmological constant $\Lambda$, and the total energy density, respectively.
The scale factor at present for the
background spacetime is set to $a_0=1$.
In what follows, we describe variables in
the inner patch by putting tilde ''$\sim$'' on top
of the variables and we consider only flat FRW universes
with dusts and a cosmological constant $\Lambda$.

First, we calculate matter density contrast $\delta_m$ of the inner patch. 
Initially ($z_i \gg 1$), we assume that  
the fluctuation of the matter perturbation $\delta_{mi}$ 
is so small that $\T{a}_i \approx a_i, \T{H}_i \approx
H_i$. Then, the Friedmann equation (\ref{eq:1}) yields,
\BE
\f{\T{H}^2}{H_0^2}=\f{\Omega_{m,0}(1+\delta_{mi})}{\T{a}^3}+\Omega_{\Lambda,0}
-\f{\delta_{mi} \Omega_{m,0}}{\T{a}^2 \T{a}_i}. \label{eq:2}
\EE
In terms of physical radius of the patch $\T{R}=\T{a} r$, where
$r$ is the comoving radius measured in the background spacetime, 
equation (\ref{eq:2}) can be written as
\BEA
\biggl (\f{d \T{R}}{d t} \biggr )^2\!
&=&\!
H_0^2\biggl[-\Omega_{m,0}(1+z_i)^3
\T{R}_i^2\delta_{mi}+\Omega_{m,0}(1+z_i)^3
\nonumber
\\
&\times&
(1+\delta_{mi})\f{\T{R}_i^3}{\T{R}} +
\Omega_{\Lambda,0}\T{R}^2  \biggr], \label{eq:3}
\EEA   
where $t$ is the cosmological time. 
The matter density contrast $\delta_m$ can be 
written as a function of a ratio of the present and the 
initial comoving radius $\eta \equiv r/r_i$ as
$\delta_m=\eta^{-3}-1$. From equation (\ref{eq:3}), $\eta$ as a 
function of redshift $z$ is given by solving 
\BEA
\biggl (\f{d \eta}{d z} \biggr )\!
&=&\!
-\bigl[-\Omega_{m,0}(1+z_i)\delta_{mi}
+\Omega_{m,0}(\eta/(1+z))^{-1}
\nonumber
\\
&+&\Omega_{\Lambda,0}(\eta/(1+z))^2
\bigr]^{1/2}[\Omega_{m,0}(1+z)^3+\Omega_{\Lambda,0}]^{-1/2}
\nonumber
\\
&+&\f{\eta}{1+z}. \label{eq:4}
\EEA
Note that right-hand-side in equation (\ref{eq:4}) does not depend on $\T{R}_i$.
From numerical integration of equation(\ref{eq:4}), 
the matter density contrast $\delta_m$ 
as a function of redshift $z$ is obtained 
by setting initial density contrast $\delta_{mi}=\delta_m(z_i)$. 

The linearly perturbed 
matter density contrast $\delta_m^L$ in the FRW background spacetime
is given by
\BE
\delta_m^L(z)=\f{3 \, \delta_{mi} H(z)}{5}\int_z^{\infty}
du \f{u+1}{H^3(1/u-1)},
\EE
where $H^2(z)=\Omega_{m,0}(1+z)^3+\Omega_{\Lambda,0}$ (Heath 1977).
Constant $3/5$ comes from our choice of initial condition that the 
peculiar velocity inside the initial patch is zero.
The relation between $\delta_m$ and $\delta_m^L$ is shown in figure 1.
\begin{figure}[t]
 \begin{tabular}{cc}
  \begin{minipage}{0.43\hsize}
   \begin{center}
    \includegraphics[width=7cm,clip]{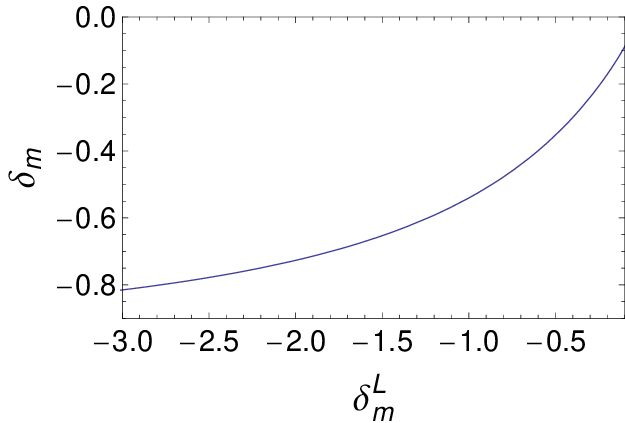}
   \end{center}
  \end{minipage}
  \begin{minipage}{0.6\hsize}
   \begin{center} 
    \includegraphics[width=6.6cm,clip]{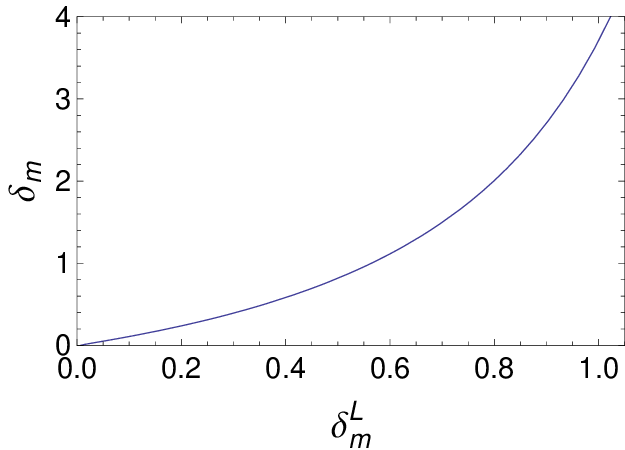}
    \end{center}
  \end{minipage}
 \end{tabular}
\vspace{0cm}
 \caption{Density contrast $\delta_m$ as a function
 of linear density contrast $\delta_m^L$ for voids (left) and clusters (right). }
\end{figure} 
Because the relation does not change so much 
even if one varies the cosmological parameters of the background
spacetime, non-linear isolated homogeneous spherical patches can be solely
calculated from corresponding linear perturbations (Friedmann \& Piran
2001). It should be noted, 
however, that the relation is valid only if $\delta_m \gtrsim
\delta_{v,cut}=-0.8$ because of shell-crossing (Furlanetto \& Piran 2006).

Next, we calculate the Hubble parameter contrast $\delta_H=\T{H}/H-1$ in 
non-linear regime. 
Plugging $\eta=R(1+z)/(R_i(1+z_i))$ into 
equation (\ref{eq:2}), we have
\BEA
&&\T{H}^2(\Omega_{m,0}(1+z)^3+\Omega_{\Lambda,0} )
\nonumber
\\
&=&H^2(\Omega_{m,0}(1+\delta_{mi})(1+z)^3 \eta^{-3}+\Omega_{\Lambda,0}
\nonumber
\\
&-& \delta_{mi} \Omega_{m,0}(1+z_i)(1+z)^2\eta^{-2}),
\EEA  
where $\eta$ is given by solving equation (\ref{eq:4}). From equation
(3) and (12), one can estimate the non-linear parameter as
\BE
\epsilon=\delta_{m}^{-1}\Omega_{m,0}(1+1/f(w))/2 -\delta_m^{-1}(\T{H}/H-1).
\EE

\begin{figure}[t]
 \begin{tabular}{cc}
  \begin{minipage}{0.41\hsize}
   \begin{center}
    \includegraphics[width=7cm,clip]{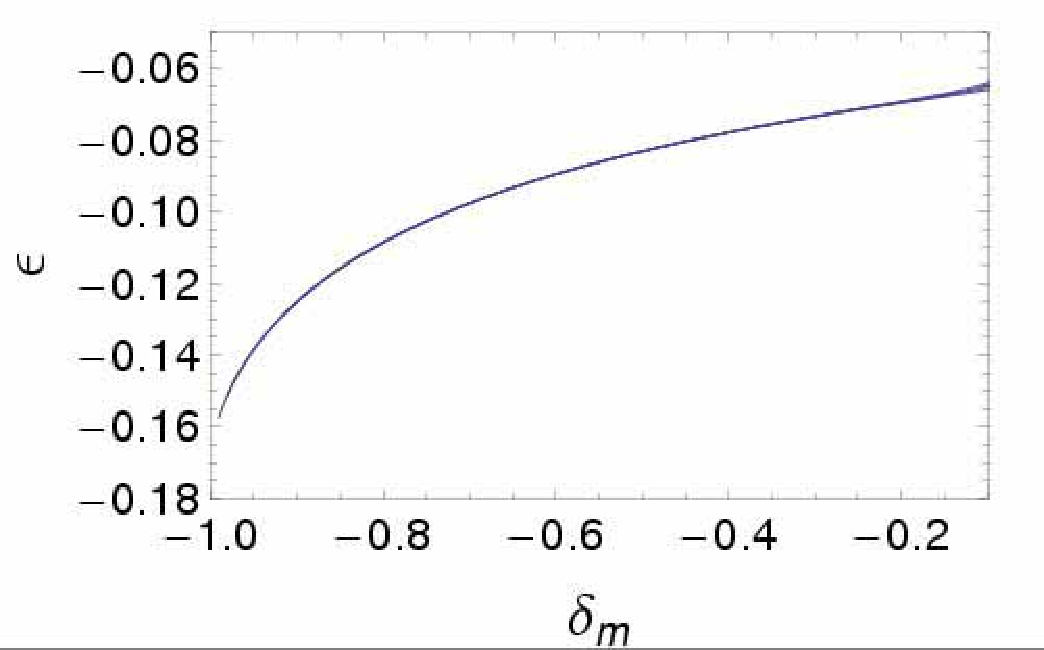}
   \end{center}
  \end{minipage}
  \begin{minipage}{0.65\hsize}
   \begin{center}
    \includegraphics[width=7.8cm,clip]{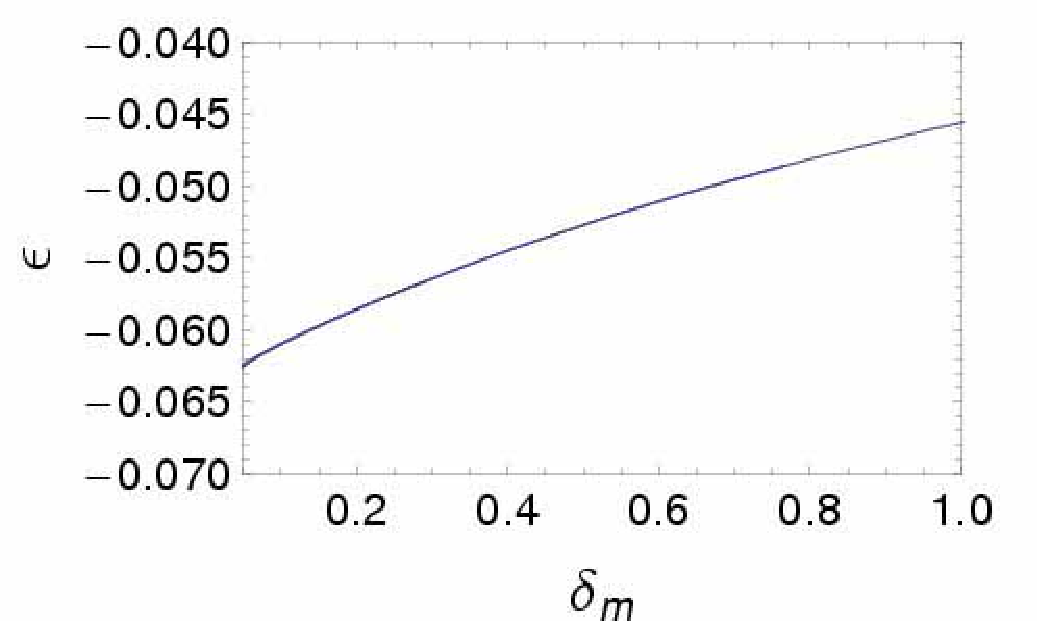}
    \end{center}
  \end{minipage}
 \end{tabular}
\vspace{0cm}
 \caption{\label{fig2} Non-linear parameter $\epsilon$ as a function of
density contrast $\delta_m$ in the EdS model. }
\end{figure} 
\begin{figure}[t]
 \begin{tabular}{cc}
  \begin{minipage}{0.45\hsize}
   \begin{center}
    \includegraphics[width=7cm,clip]{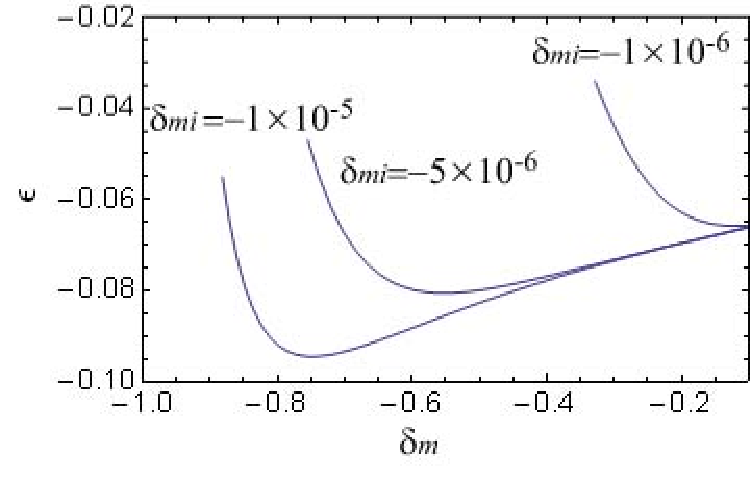}
   \end{center}
  \end{minipage}
\hspace{-0.6cm}
  \begin{minipage}{0.6\hsize}
   \begin{center}
    \includegraphics[width=7.4cm,clip]{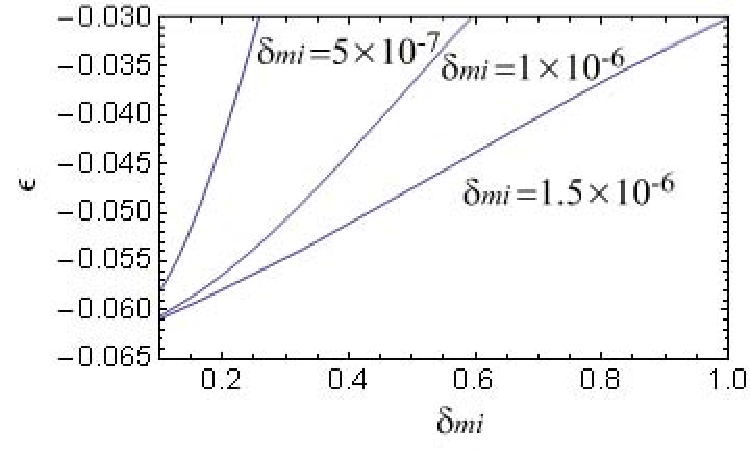}
    \end{center}
  \end{minipage}
 \end{tabular}
\vspace{0cm}
 \caption{Non-linear parameter $\epsilon$ as a function of
density contrast $\delta_m$ in the flat-$\Lambda$ model with 
$\Omega_{m,0}=0.26$.}
\end{figure} 
\subsection{Effect of non-linear dynamics}
In previous section, we have seen that the non-linear density contrast
$\delta_m$ for a spherically symmetric homogeneous patch 
can be written in terms of corresponding 
linear density contrast $\delta^L_m$.
In order to calculate the ISW effect, we need to estimate the Hubble 
parameter contrast $\delta_H$ and the peculiar velocity $\tilde{v}$ 
of the wall. The non-linear corrections to $\delta_H$ and $\tilde{v}$ 
can be characterized by two parameters $\epsilon$ and $\nu$, respectively.

First, we consider the effect of non-linear correction to the Hubble contrast 
$\delta_H$.  
As one can see in figure 2 and 3,  
$\epsilon$ is always negative and the amplitude is 
$|\epsilon|<0.16 $ for $|\delta_m|<1.0$. This represents a slight enhancement 
in the expansion speed within the inner patch due to non-linearity.
In low-density universes ($\Omega_{m,0}<1$),
$|\epsilon|$ is smaller than that in high-density universes. 
In the Einstein-de Sitter (EdS) universe, $\epsilon$ depends only on
$\delta_m$ (figure 2). In contrast, in low-density universes,  $\epsilon$
depends on the amplitude of the initial epoch as well (figure 3).
This is because the expansion speed inside the patch 
is suppressed when the energy
component of the background 
universe is dominated by a cosmological constant $\Lambda$. 
We have found that the non-linear contribution to $\delta_H$ 
is less than 10 per cent
for $|\delta_m|<0.2$ and $\Omega_m>0.26$.

Second, we consider the effect of non-linear correction to peculiar
velocity of the wall. In the thin-shell limit, the motion of the 
spherically symmetric wall can be obtained by numerically solving a set of 
ordinary differential equations using Israel's method(Israel 1966, Maeda
\& Sato 1983).
If the inner region and the outer region are described by  
the FLRW spacetime, the fitting formula for the peculiar velocity
of an expanding wall normalized by the background Hubble expansion 
can be written as (Maeda, Sakai, \& Triay 2010)
\BE
\tilde{v}=\f{\Omega_{m}^{0.56}}{6}(|\delta_m|+0.1 
\delta_m^2+0.07|\delta_m|^3), \label{eq:tildev}
\EE
for $\Omega_{m}+\Omega_{\Lambda}=1$. We have confirmed that 
the accuracy of the fitting formula is within one percent
for $0<\Omega_{m} \le 1$ and $|\delta_m|<1$ using numerically computed values. 
From equation (\ref{eq:tildev}), we find that 
the contribution of non-linear effect is less than 5 percent 
for $|\delta_m|<0.3$. Thus 
an approximation $\delta_H \propto \delta_m \propto \tilde{v}$
can be validated in the quasi-linear regime. 

 \subsection{Effect of non-linearity and inhomogeneity on the ISW signal}
 
\begin{figure}[tbp]
\begin{center}
   \hspace{-0.4cm}
    \includegraphics[width=17cm,clip]{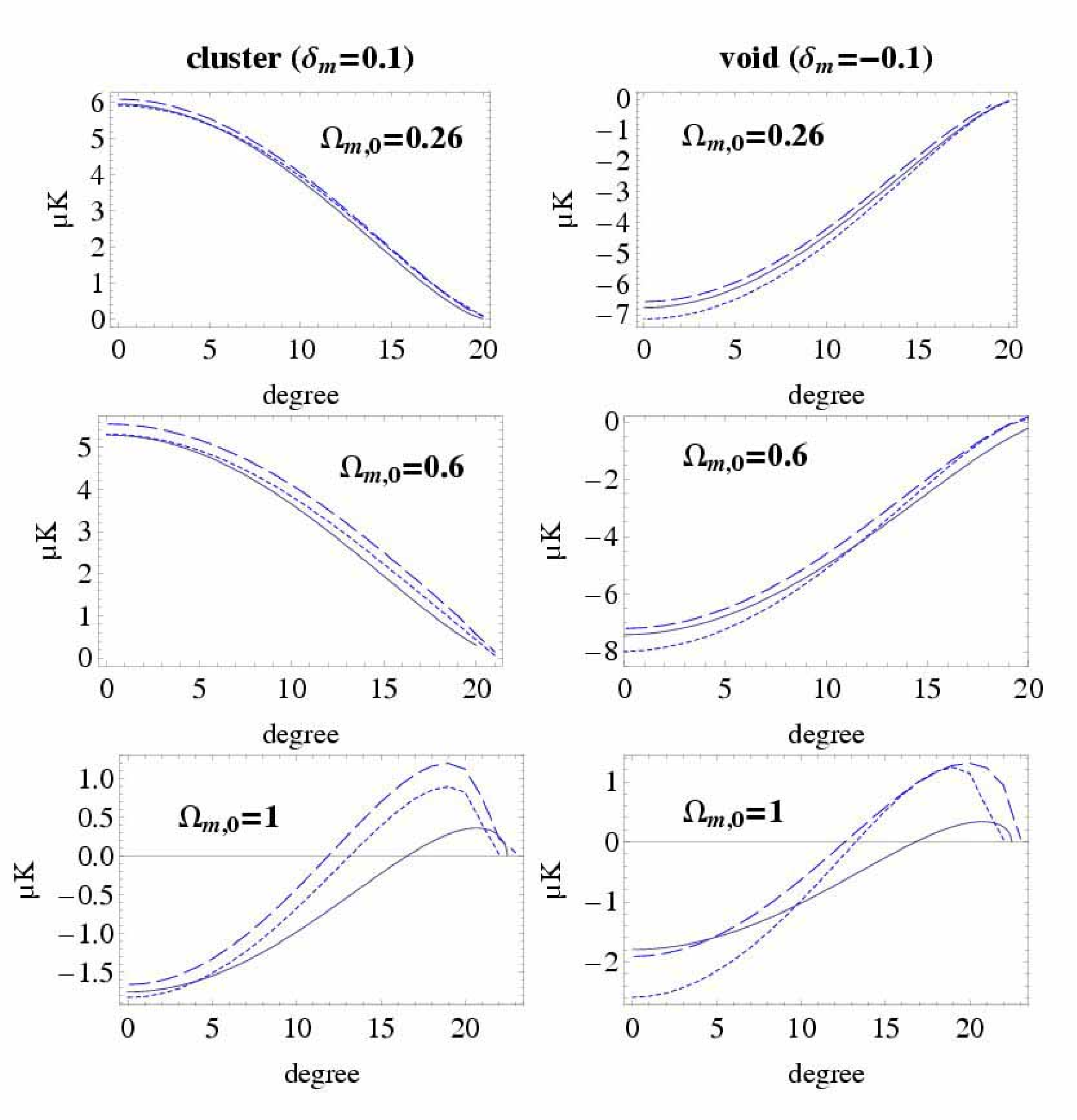}
   \label{figure4}
\end{center}
\vspace{-0.32cm}
 \caption{Temperature fluctuation $\Delta T$ due to a compensated
 cluster (left panels)/a compensated void (right panels) centered at redshift
 $z=0.2$ as a function of angular radius from the center. We used
 three methods for deriving the ISW
 signal:
the thin-shell approximation (thick curve), the second order perturbation theory
 (dotted curve) and the LTB solution (dashed curve). 
We set the comoving (outer) radius of the cluster/void $r_v=200\,h^{-1}$Mpc for the thin-shell approximation, $r_{out}=210\,h^{-1}$Mpc for
the second order and the LTB calculations with wall width 
about tenth of the outer radius.  The density parameter is  
$\Omega_{m0}=0.26$. 
For detail, see the text.}
\end{figure} 

\begin{figure}[t]
\begin{center}
   \hspace{-1cm}
    \includegraphics[width=17cm,clip]{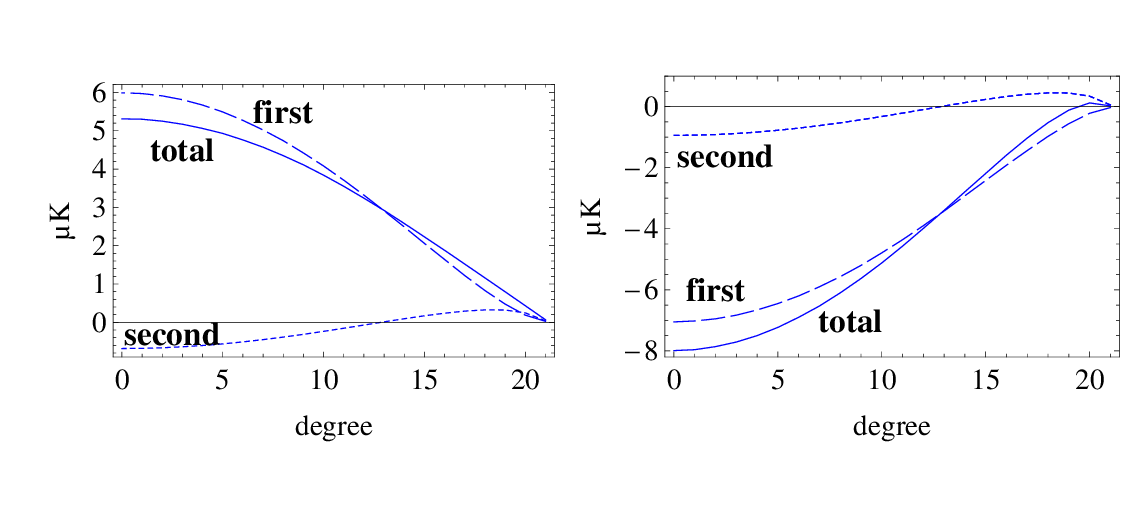}
   \label{figure5}
\end{center}
\vspace{-1cm}
  \caption{The first and the second order contributions to temperature fluctuation $\Delta
 T$ due to a compensating cluster (left)/void (right) centered at redshift $z=0.2$ as a function of an angular
 radius from the center. We set the outer comoving radius 
$r_{out}=210\,h^{-1}$Mpc, the inner comoving radius $r_{in}=0.93\, r_{out}$, the density parameter $\Omega_{m0}=0.6$, and the
 density contrast at the center of the cluster/void $|\delta_m(z=0.2)|=0.1$. }
\end{figure} 
\begin{figure}[]
\hspace{-0.5cm}
\begin{center}
    \includegraphics[width=17cm,clip]{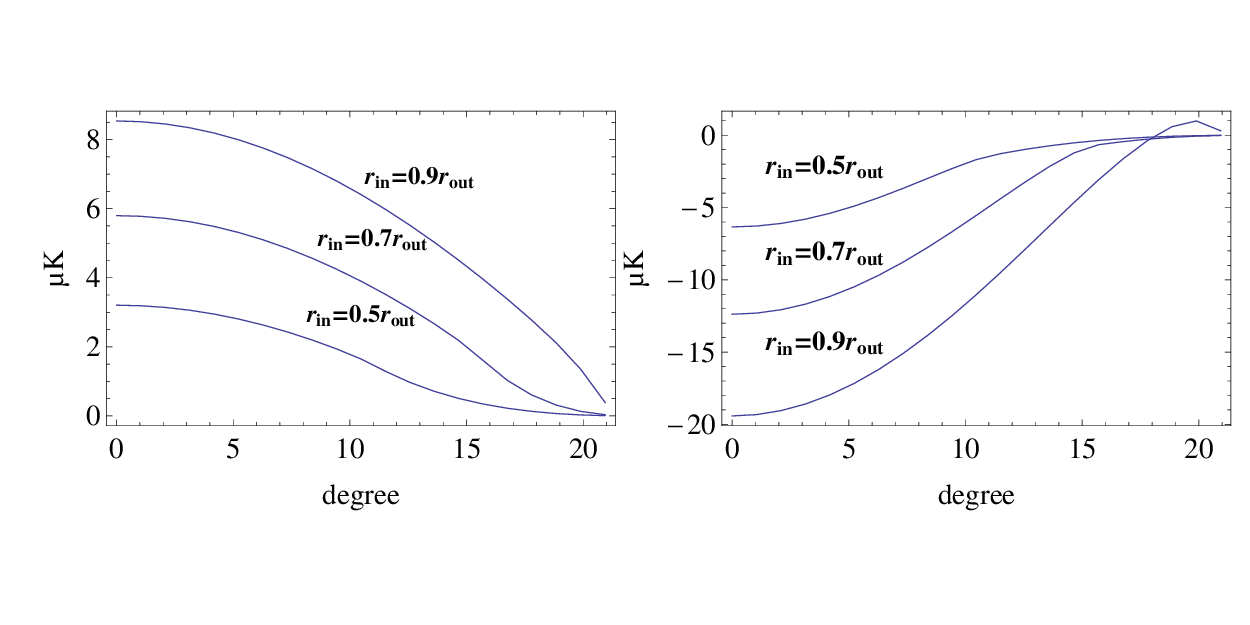}
   \label{figure6}
\end{center}
\vspace{-0.9cm}
  \caption{The effect of thickness of the wall for $\Omega_{m0}=0.6$ 
$r_1=210\,h^{-1}$Mpc, and
 $\delta_{m}(z=0.2)=0.2$ (left) and $\delta_{m}(z=0.2)=-0.2$ (right)
for top-hat type density perturbations (see appendix A). }
\end{figure} 
In literature, the thin-shell approximation has been often used to
describe almost empty voids
with $\delta\sim -1$ (Maeda \& Sato 1983). 
In quasi-linear regime, however, we also need to
consider the effect of thickness of the wall and inhomogeneity 
of the matter distribution because quasi-linear voids are not  
in the asymptotic regime. Non-linearity of the wall may significantly affect 
the CMB photons that pass through it.  Moreover,
it seems not realistic to apply the thin-shell 
approximation to spherically symmetric
clusters since the mass of the wall cannot be negative. 

In order to estimate the validity of the thin-shell approximation, 
we have compared the ISW signal with those obtained by using 
second order perturbation theory (Tomita \& Inoue 2008) and
by using the Lemaitre-Tolman-Bondi (LTB) solution (Sakai \& Inoue 2008), 
which yields exact results without recourse to the 
cosmological Newtonian approximation. 
We have assumed 
top-hat type matter distribution (for linear matter perturbation) 
for void/cluster for calculation using second order perturbation theory
and a smooth distribution specified by a certain polynomial function
for calculation using the LTB solution. The voids/clusters are 
assumed to be compensated so that the gravitational potential outside 
the cut-off radius ($r_1$ for the perturbative analysis and $r_{out}$
for the LTB-based analysis) is constant. 
For detail, see appendices A and B.     

As an example, we have computed temperature fluctuations $\Delta T$
generated from a compensated
void/cluster using the three types of method.
The density contrast, the comoving radius and the 
redshift of the center of a void/cluster are set to $|\delta_m|=0.1$, 
$r_v=200\,h^{-1}$Mpc, $r_1=r_{out}=210\,h^{-1}$Mpc, and $z=0.2$,
respectively.
The width of the wall is assumed to be 1/10 of the cut-off radius.
As one can see in figure 4, the three methods agree well 
for low density universes in which the linear effect is dominant. 
In contrast, the discrepancy becomes apparent 
for high density universes in which the non-linear effect is dominant.
This discrepancy is partially due to a slight difference in the assumed
density profile (top-hat type for the perturbative
analysis, polynomial type for the LTB).  
In order to demonstrate the role of non-linearity, we have plotted
first order (linear ISW effect) and second order (RS effect minus linear
ISW effect) contributions to the ISW signal (figure 5). The first order
effect makes the CMB temperature negative(positive) for a void(cluster)
but the second order effect makes the CMB temperature negative  
near the center and positive near the boundary regardless
of the sign of the density contrast. As a result, 
the amplitude of temperature fluctuation 
for a void(cluster) is enhanced(suppressed) 
in the direction near the center but it is
 suppressed(enhanced) in the direction 
near the boundary.   These non-linear effects become
much apparent for models with higher background density 
because the linear ISW effect becomes less effective. 
However, if we take into account of the thickness
of the wall, these non-linear effects can be less conspicuous since the 
amplitude of the gravitatinal potential becomes smaller for a fixed
outer radius (figure 6).   
In the $\Lambda$-dominated universe, a quasi-linear compensated void 
can be recognized as a cold spot surrounded by a very weak hot ring, whereas
a quasi-linear compensated cluster can be recognized as a hot spot possibly
with a dip at the center of it. In the EdS universe, either a
compensated quasi-linear void or a cluster can be identified as a cold spot
surrounded by a hot ring.  
\subsection{ISW effect from prominent quasi-linear structures}
In order to fully utilize information of the three dimensional 
distribution, we consider a temperature anisotropy $\it{\Delta} T/T$ 
obtained from stacked images on the CMB sky that
corresponds to most prominent voids/clusters in a galaxy catalog.  
First, we fix an angular radius $\theta_{out}$ 
of a circular region on the CMB sky that will be used in the stacking
analysis. Then, the corresponding smoothing scale $r_s$ in comoving coordinate 
for the corresponding fluctuation at $z$ is  
$r_s=(1+z)D_A(z) \theta_{out}$, where $D_A(z)$ is the angular 
diameter distance to the galaxy. The corresponding 
initial smoothing scale is $r^i_s=\eta^{-1}r_s$. 

We assume that 
the probability distribution function (PDF) of linear density 
contrast $\delta_m^L$ at redshift $z$ is given by a Gaussian distribution function,  
\BE
P^L(\delta_m^L;\sigma(r^i_s,z))=\f{1}{\sqrt{2 \pi}\sigma(r^i_s,z)}\exp 
\biggl[-\f{(\delta_m^L)^2}{2 \sigma^2(r^i_s,z)} \biggr] \label{eq:5}
\EE
where $\sigma^2(r^i_s,z)$ is the variance of the linearly extrapolated 
density contrast at redshift $z$ smoothed by a spherically symmetric 
top-hat type window function with an initial comoving radius $r^i_s$.
Note that $\sigma(r^i_s,z)$ depends on cosmological parameters such as 
$\Omega_{m,0}, \sigma_8$ and the spectrum index $n$. 
Then, the PDF of non-linear density contrast of the inner patch $\delta_m$
is given by
\BE
P^{NL}(\delta_m;\sigma(r^i_s,z))=\alpha P^L(\delta_m^L(\delta_m);\sigma(r^i_s,z))\f{d
\delta_m^L}{d \delta_m}, \label{eq:6}
\EE
where $\alpha$ is a constant that normalizes the PDF.
\begin{figure}[tbpm]
\hspace{-0.5cm}
\begin{center}
    \includegraphics[width=9cm,clip]{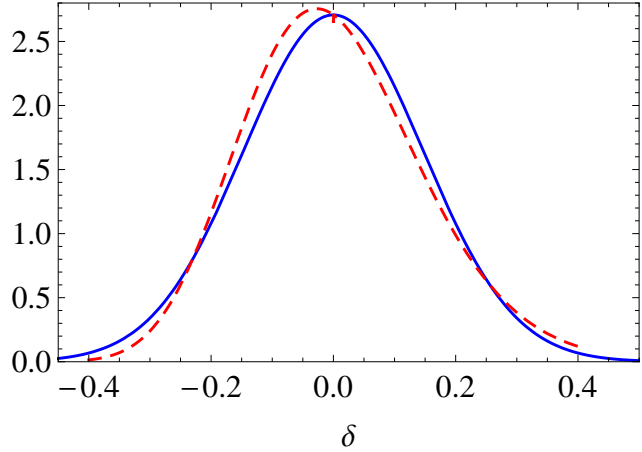}
   \label{figure6}
\end{center}
\vspace{-0.9cm}
  \caption{Effect of non-linearity: the PDF of $\delta_m$ (dashed curve)
and that of $\delta^L_m$ (full curve) for a smoothing scale $r_s=50h^{-1}$Mpc at $z=0.5$. }
\end{figure} 

As shown in figure 7, the PDF of $\delta_m$ is positively skewed in comparison with the PDF of $\delta_m^L$ because of non-linearity.  
For a sample region at redshift $z$ with a total comoving volume
$V$, the total number of voids or clusters with a radius $r_s$
is approximately $N_t \approx 3V/(4 \pi 
r^3_s)$. In what follows we assume that the number of
prominent voids/clusters ($N_v/N_c$) determines the corresponding 
threshold of density contrast $\delta_{m,th}(z)$, which is given by
\BE
N_v/N_t=\int_{\delta_{v,cut}}^{\delta_{m,th}}
P^{NL}(\delta_m;\sigma(r^i_s,z))d \delta_m, \label{eq:7}
\EE
and 
\BE
N_c/N_t=\int_{\delta_{m,th}}^{\delta_{c,cut}} 
P^{NL}(\delta_m;\sigma(r^i_s,z))d \delta_m, \label{eq:8}
\EE
respectively. From equations (\ref{eq:dToverTthin}), (\ref{eq:7}), and 
(\ref{eq:8})  the mean temperature fluctuation 
within an angular radius $\theta_{out}$ for a
stacked $N_v$ or $N_c$ images corresponding to prominent quasi-linear
voids/clusters at redshift $\sim z$ can be approximately written as
\BE
\langle \!\it{\Delta T}\rangle\!=\f{\int\!\!\! \int
W(\theta;\theta_{in})\it{\Delta T}(\delta_m, \psi(\theta))
P^{NL}(\delta_m;\sigma)d^{2}{\boldsymbol{\theta}} d\delta_m}
{ \pi \theta_{th}^2 \int P^{NL}(\delta_m;\sigma)d\delta_m   },
\label{eq:stackdT}
\EE
where $0 \le \theta \le \theta_{out}$, $W(\theta;\theta_{in})$ is a
compensating window function 
that satisfies
\BE
\int_0^{\theta_{in}} 2 \pi \theta  W(\theta;\theta_{in}) 
d\theta =-\int_{\theta_{in}}^{\theta_{out}} 2 \pi \theta 
W(\theta;\theta_{in}) d\theta
\EE
and 
\BE
\psi(\theta)=\theta + \arcsin
\biggl[h^{-1} \cos\theta \sin\theta - \sqrt{\sin^2\!\theta - 
h^{-2} \sin^4 \!\theta} \biggr],
\EE
where $h=r_s(1+z)/D_A(z)$. 

\section{Application to observations}

\subsection{SDSS-WMAP cross correlation}
A cross correlation 
analysis using a stacked image built by
averaging the CMB surrounding distinct voids/clusters
has been done by Granett et al. (2008).
They have used 1.1 million Luminous Red
Galaxies (LRGs) from the SDSS catalog covering 7500
square degrees. The range of redshifts 
of the LRGs is $0.4<z<0.75$, with a median of 
$\sim0.5$. The total volume is $\sim 5\,h^{-3}
\textrm{Gpc}^3$. They used so-called the ZOBOV (ZOnes Bordering On Voidness;
Neyrinck 2008) algorithm to find 
supervoids and superclusters in the LRG catalog
and made a stacked image from an inversely 
variance weighted WMAP 5-year (Q,V, and W) map. 
In order to reduce contribution from CMB fluctuations
on scales larger than the objects, they used a top-hat type 
compensating filter 
\BE
W_{th}(\theta;\theta_{in})=
\left\{ 
\begin{array}{ll}
  1 & (\theta<\theta_{in}) \\
\!  -1 & (\theta_{in}\le 
\theta \le \theta_{out}),  \\
\end{array}
\right.
\EE 
where $\theta_{out}=\cos^{-1}{(2 \cos\theta_{in}-1)}$.

First, using the developed tools based on thin-shell approximation
and homogeneous collapse model in section 2, 
we estimate the expected amplitude of the ISW signal for prominent 
structures in a 
concordant $\Lambda$CDM model with Gaussian primordial fluctuations
and compare with the observed values obtained from the SDSS-LRG
catalog. The number of most distinct 
voids or clusters $N$ and the cut-off radius $\theta_{in}$ are 
chosen as free parameters. At redshift $z=0.5$
, the mean density contrast filtered by a top-hat type function
with radius $r=130\,h^{-1}$Mpc corresponding to
$\theta_{out}\sim 5.6^\circ$ is just $\langle \delta_m^L\rangle=0.046$
and the background density parameter is $\Omega_m(z=0.5)=0.54$.
Because the influence of non-linear ISW effect is weaker than that of the 
linear ISW effect in this setting, we expect that details
of non-linear calculations will not much affect the result.
In what follows, we 
use an approximation $\delta_H \propto
\delta_m \propto \tilde{v}$, where $\delta_m$ is
determined from the homogeneous collapse model
in section 2.  
 
As shown in table 1, it turned out that the expected 
values of the ISW(Rees-Sciama) signal are typically 
of the order of $O(10^{-7})$K. As expected, the amplitude
gets larger as the number of stacked image decreases, and the amplitude
for voids systematically becomes larger than those for clusters by
$5-10$ percent(Tomita \& Inoue 2008; Sakai \& Inoue 2008). 
On the other hand, the order of the observed
amplitudes are extremely large as $O(10^{-6})$K. It turns out that 
the discrepancy remains at $3-4\,\sigma$ level for $N=30$ and $N=50$. 

Second, we reconstruct the mean density profile from
the observed ISW signal for the SDSS-LRG catalog 
using our LTB model.  From figure 7, one can notice a
hot ring around a cold spot for the stacked image of voids and
a dip at the center of a hot spot for the stacked image of
clusters. Although the amplitude of the hot-ring cannot be
reproduced well, the observed
dip at the center of the hot spot can be qualitatively reproduced 
in our LTB models. We have found that the dip at the center of a compensated 
cluster can be generated only if the linear ISW effect balances  
the non-linear ISW effect in a limited parameter region. 
Thus the observed features in stacked images 
strongly imply that the corresponding super-structures are 
not linear but quasi-linear or non-linear objects. The density 
fluctuations which are necessary to produce the observed ISW signals 
are found to be tremendously large.  In figure 8, we plot the 
ISW signal from a compensated cluster with $\delta_m \sim 7\,\sigma$
and that from a compensated void with $|\delta_m|\sim 10\,\sigma$ at 
$z=10$ (see the radial density profiles at figure A1 in appendix A). 
Even for these very rare objects, the 
amplitudes of ISW signals in our LTB models are much
smaller than the observed ones. In fact, 
the mean temperatures for a compensating filter $\theta_{in}=4 ^\circ$
are $3.6\mu \textrm{K}(1.4\, \sigma)$ and $-3.1\mu \textrm{K}
(2.6\,\sigma)$ for the cluster and the void, respectively. 
On the other hand, the probability 
of generating these fluctuations is as extremely small as $10^{-12}$
in standard inflationary models that predict primordial gaussianity. 

Thus, the observed large ISW signals 
for the stacked image strongly suggest a presence of super-structures
on scales O(100$\,h^{-1}$Mpc) with anomalously large density contrast 
O(0.1) which can not be produced in the concordant LCDM model. 
\begin{deluxetable}{l l l l}
\label{tab1}
\tablecolumns{4}
\tablecaption{Expected and observed amplitude of mean temperature for a compensating filter 
$\theta_{in}=4^\circ$}
\tablehead{$N$ & cluster ($\mu K$)& void ($\mu K$)& average($\mu K$)}  
\startdata
1 & 0.98 & -1.2 & 1.07\\
5 & 0.82& -0.94  & 0.88 \\
10& 0.73 &-0.83 & 0.78\\
30 & 0.57 &-0.64 & 0.61 (11.1 $\pm2.8$)\tablenotemark{a} \\
50 & 0.48 (7.9 $\pm3.1$)\tablenotemark{a} &-0.53 (-11.3 $\pm3.1$)\tablenotemark{a} & 0.51 (9.6 $\pm2.2$)\tablenotemark{a}\\
70 & 0.42 & -0.46 & 0.42 (5.4 $\pm1.9$)\tablenotemark{a}
\enddata
\tablenotetext{a}{Taken from Granett et al. (2008). }
\end{deluxetable}
\begin{figure}[t]
\begin{center}
   \hspace{-1cm}
    \includegraphics[width=16cm,clip]{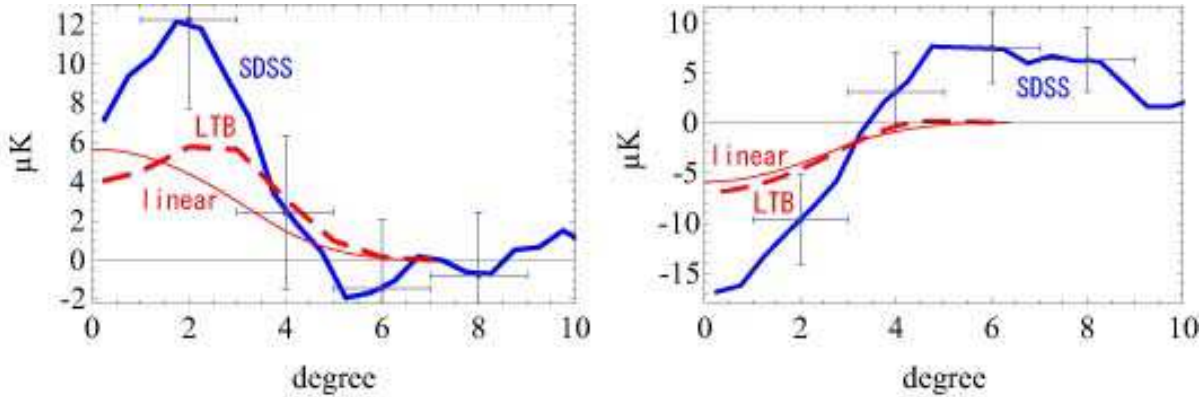}
\end{center}  
\vspace{0cm}
 \caption{\label{fig7}  Observed mean CMB temperature radial profiles 
for a stack image of 50 clusters (left) and that of 50 voids (right)
(thick full curves) in the SDSS-LRG catalog (Granett et al 2008) and the
 corresponding theoretical
radial profiles for compensated spherical clusters (left)/void (right)
based on the LTB solution (dashed curves) and the 
linear calculation (thin full curves) 
with a top-hat type compensated density distribution corresponding to
 the LTB solutions.   
The LTB parameters for the set of clusters are
$(r_{out},r_{in},\delta_m,z_c)=(0.055H_0^{-1},0.022H_0^{-1},1.0,0.5)$
and those for the set of voids are
 $(0.050H_0^{-1},0.015H_0^{-1},-0.55,0.5)$. The effective
 radii of the inner patch at a redshift $z=10$ and the mean filtered 
temperature with $\theta_{in}=4^\circ$ are $(0.029H_0^{-1}, 3.6
 \mu\textrm{K})$ (cluster) 
and $(0.023H_0^{-1}, -3.1 \mu\textrm{K})$ (void), respectively.  
Evolution of the density profile in the LTB models are shown in appendix B.  
$1\sigma $ error bars are obtained from 1000 Monte Carlo
simulations on the WMAP7 Q+V+W map smoothed at $1^\circ$ scale (see section
 3.3).  }
\end{figure} 
\subsection{2MASS-WMAP cross correlation}

Francis \& Peacock (2009) estimated the local
density field in redshift shells using photometric
redshifts for the 2MASS galaxy catalog. They reconstructed
the CMB anisotropies due to the ISW effect from
the local density field $\delta_m$. They approximated the bias 
in each redshift shell by a linear bias relation $\delta_g=b\,\delta_m$
and assumed that the bias is independent of scale and redshift in each
shell. In order to obtain the bias parameter $b$, a maximum likelihood analysis
of the galaxy catalog was performed.  

There are two prominent 
spots in the reconstructed CMB anisotropy. One is 
a hot spot due to a supercluster around the Shapley concentration
at redshifts $0.1 \le z \le 0.2$. Another one is a cold spot due to
a supervoid at redshifts $0.2 \le z \le 0.3$ in the direction to 
$(l,b)\sim (0,-30^\circ)$. The angular radii of both structures are 
$\theta=20^\circ-30^\circ$. The temperatures near the center of
both structures are $\sim 20\,\mu$K. The position of the 
supervoid is very close to the one predicted in Inoue \& Silk (2007), $(l,b)\sim (330^\circ, -30^\circ)$.  

Based on developed methods in section 2, we have estimated the
expected density contrast $\delta_m$ and the corresponding temperature profile 
(figure 9) due to a most prominent object in the shell.  We have assumed
the same cosmological parameters and primordial gaussianity as those
discussed in section 3.    
In order to compute the temperature profile, we have used a homogeneous thin-shell 
model. As shown in table 2, the observed density contrasts 
are larger by 4-7 times the expected values. If the  
comoving radius of the structure is $\sim 200\,h^{-1}$Mpc,
the absolute value of the density contrast should be $|\delta_m|=0.2-0.3$,
which implies a presence of anomalous quasi-linear super-structures. 
Our result is consistent 
with the power spectrum analysis in Francis \& Peacock (2009)
where a noticeable excess of the observed power at low multipoles $2 \le l \le4$
was reported.

\begin{deluxetable}{l l l l l l}
\label{tab2}
\tablecolumns{6}
\tablecaption{Expected and observed density contrast for super-structures
in 2MASS galaxy catalog}
\tablehead{radius\tablenotemark{a} & expected & observed & radius\tablenotemark{a}  & expected & observed } 
\startdata
 230 & 0.037 & 0.20 & 370 & -0.013 & -0.049 \\
150 & 0.094 & 0.69  & 250 &  -0.037  & -0.15 \\
\enddata
\tablenotetext{a}{The unit of the radii is $h^{-1}$ Mpc.  }
\end{deluxetable}
\begin{figure}[t]
 \begin{tabular}{cc}
\hspace{-0.7cm}
  \begin{minipage}{0.5\hsize}
\vspace{0.1cm}
   \begin{center}
    \includegraphics[width=6.9cm,clip]{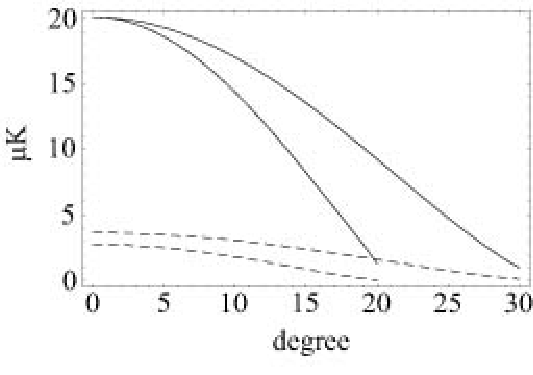}
   \end{center}
  \end{minipage}
  \begin{minipage}{0.5\hsize}
   \begin{center}
    \vspace{-0.4cm}
    \hspace{-0.3cm}
    \includegraphics[width=7.1cm,clip]{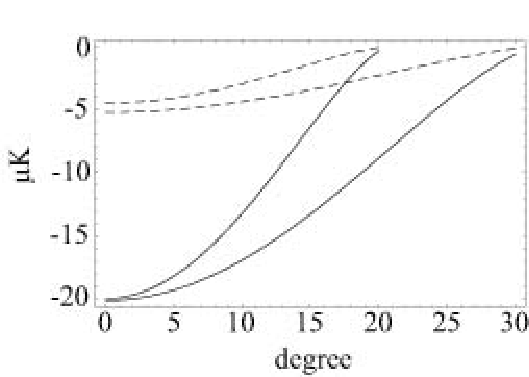}
    \end{center}
  \end{minipage}
 \end{tabular}
\vspace{0cm}
 \caption{\label{fig2} Left: Theoretically modeled temperature profiles for an observed 
(full curves) and an expected (dotted curves) supercluster in the 2MASS catalog.
Right: modeled temperature profiles for an observed 
(full curves) and an expected supervoid (dotted curves).
These superstructures reside at redshifts $z=0.1-0.3$. 
We have plotted two possible profiles for each structure since
there is an ambiguity in the angular size due to errors ($\Delta z\sim
 0.03$) in photometric redshifts in the 2MASS galaxy catalog (Francis \&
 Peacock 2009). }
\end{figure} 
\vspace{0.5cm}
\subsection{The CMB cold spot}
The most striking CMB anomaly is the presence of 
an apparent cold spot in the Wilkinson Microwave Anisotropy
Probe (WMAP) data in the Galactic southern hemisphere (Vielva et al. 
2004; Cruz et al. 2005) (see figure 10). 
The cold spot has a less than 2 per cent
probability of being generated as random gaussian fluctuations 
(Cruz et al. 2007a), if one uses spherical mexican-hat type wavelets 
as filter functions (see also Zhang \& Huterer 2010).  

Assuming that it is not a statistical artifact, a
variety of theoretical explanations have been proposed, such as
galactic foreground(Cruz et al. 2006), texture (Cruz et al. 2007b), and 
Sunyaev-Zeldovich (SZ) effect. However, these models
failed to explain other large-angle anomalies by the same
mechanism. 

Inoue \& Silk (2006,2007) proposed that the cold spot may be produced by
a supervoid at $z<1$ in the line-of-sight due to the ISW effect and  
have shown that another pair of supervoids that are tangential to 
the Shapley concentration can explain the 
alignment between the quadrupole and the octopole in the CMB. 
Subsequently, Rudnick et al. (2007) found a depression in source
counts in the NRAO VLA Sky Survey(NVSS) in the direction to 
the cold spot, although the statistical significance
has been questioned (Smith \& Huterer 2010). Recent optical observations
(Granett et al. 2009, Bremer et al. 2010), however, revealed that any 
noticeable supervoids 
at $0.35<z<1$ in the line-of-sight are ruled out. These observations
suggest that the angular size of the supervoid may be larger or smaller 
than expected and that it resides at 
low redshifts $z<0.35$ or at high redshifts $z>1$.
\begin{figure}[t]
 \begin{tabular}{cc}
  \begin{minipage}{0.45\hsize}
   \begin{center}
    \includegraphics[width=5.8cm,clip]{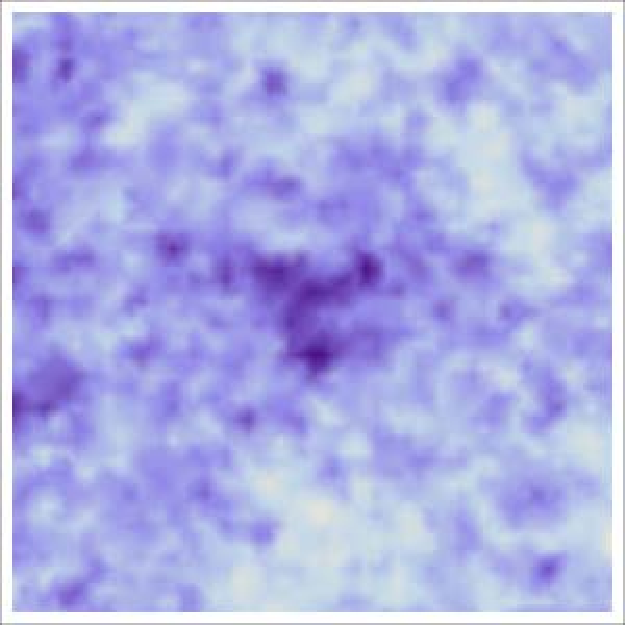}
   \end{center}
  \end{minipage}
  \begin{minipage}{0.5\hsize}
   \begin{center}
    \includegraphics[width=8cm,clip]{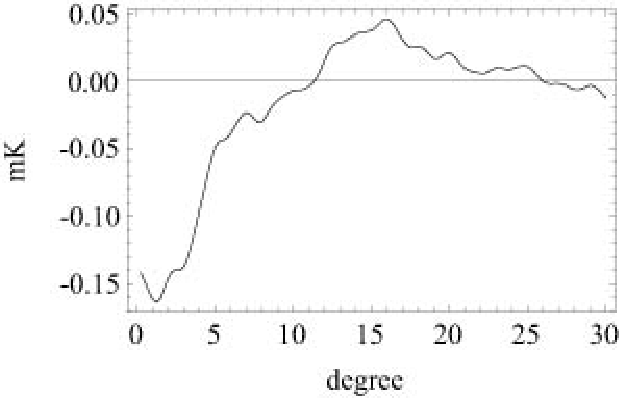}
    \end{center}
  \end{minipage}
 \end{tabular}
\vspace{0cm}
 \caption{Left: the WMAP7 ILC temperature map $(40^\circ
 \times 40^\circ$) smoothed at $1^\circ$ scale. Right: the averaged radial profile of 
the ILC map as a function of 
inclination angle $\theta$ from the center of the cold spot 
$(l,b)=(207.8^\circ, -56.3^\circ)$. 
A peak at $\theta\sim 15^\circ$ corresponds to a hot 
ring. }
\end{figure} 

In order to test such a possibility, we have calculated 
averaged temperature around the cold spot (see figure 11) 
using a spherical top-hat compensating filter $W_{th}(\theta ;\theta_{in})$.

Interestingly, we have discovered two peaks in the plot 
of the filtered mean temperature around the cold spot 
as a function of inner radius of the filter (figure 12). 
The inner and the outer peaks
are observed at $\theta_{in}=4^\circ-5^\circ$($\theta_{out}=6^\circ-7^\circ$) 
and $\theta_{in}=12^\circ-13^\circ$ ($\theta_{out}=16^\circ-17^\circ$).
The outer peak corresponds to a hot ring,  
which is visible by eyes (see figure 10).

In order to estimate the statistical significance of the peaks, 
we have used a WMAP 7-year internal linear combination (ILC) map 
smoothed at $1^\circ$ scale with a Galactic skycut $|b|<20^\circ$ 
and a combination of the Q, V, and W band frequency maps 
smoothed at $1^\circ$ scale averaged with 
weights inversely proportional to the noise variances
with a ``standard'' Galactic skycut made by the WMAP team.
In order to reduce possible residual 
contamination from the Galactic foreground, 
we further cut a region $|b|<35^\circ$ for the Q+V+W map. 
In order to estimate the errors, firstly, we generated 
1000 random positions on the ILC($|b|>20^\circ$) and
the Q+V+W($|b|>35^\circ$) maps, and then computed variances $\sigma^2$ for the filtered mean 
temperature. Second, we have calculated expected $\sigma^2$ 
for the filtered temperature 
using the angular power spectrum $C_l$ for the WMAP 
7-year data obtained by the WMAP team (see appendix A).
Note that we have computed pseudo-$C_l$'s from the $C_l$'
for each skymap.  As shown in figure 11, the observed 
standard deviations are $1 \sigma=19\sim 20 \,\mu$K 
for $\theta_{in}=4^\circ$ and $1 \sigma=14\sim 16 \,\mu$K 
for $\theta_{in}=14^\circ$. Our result for $\theta_{in}=4^\circ$ 
is roughly consistent with the values for stacked images 
in Granett et al. (2008) assuming no correlation between
voids/clusters in a particular configuration. In the Q+V+W map, a slight
suppression in $\sigma$ is observed at $\theta_{in}>8^\circ$. 
Theoretically calculated  values are
found to be systematically lager than the observed values by 
$4-16$ per cent for $4^\circ<\theta_{in}<14^\circ$.
These discrepancies represent an uncertainty due to the Galactic 
foreground emission. 
\begin{figure}[t]
 \begin{tabular}{cc}
  \begin{minipage}{0.5\hsize}
   \begin{center}
\hspace{-1cm}
    \includegraphics[width=7.3cm,clip]{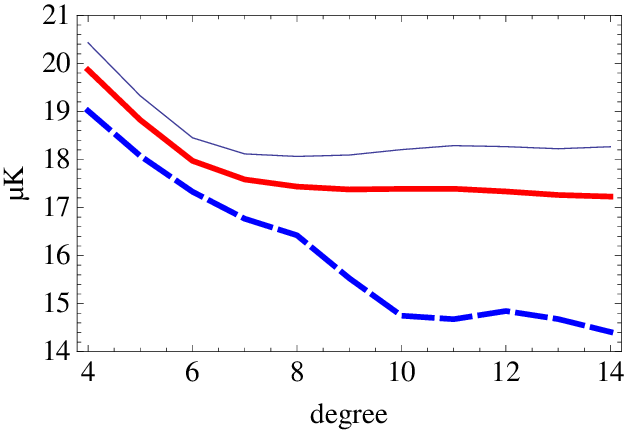}
   \end{center}
  \end{minipage}
  \begin{minipage}{0.5\hsize}
   \begin{center}
    \includegraphics[width=7.2cm,clip]
{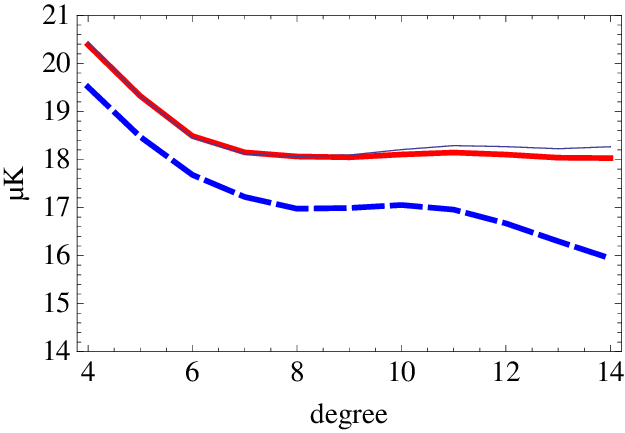}
    \end{center}
  \end{minipage}
 \end{tabular}
\vspace{0cm}
 \caption{Left: $1\sigma$ of the 
filterd mean temperature  as a function of
an inner radius $\theta_{in}$
for the Q+V+W map (dashed curve)
and that derived from 
$C_l$'s (thin curve) and from pseudo-$C_l$'s (thick curve).
Right: $1\sigma$ for the 
ILC map (dashed curve) and corresponding 
theoretical values (thin curve and thick curve) as 
shown in the left figure.}
\end{figure} 
\begin{figure}[t]
 \begin{tabular}{cc}
  \begin{minipage}{0.5\hsize}
   \begin{center}
\hspace{-1.5cm}
    \includegraphics[width=7.3cm,clip]{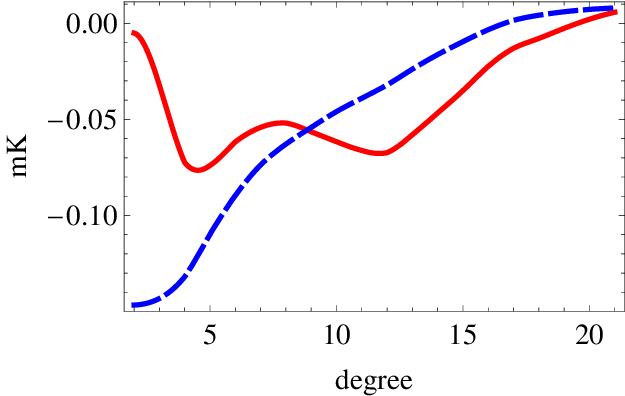}
   \end{center}
  \end{minipage}
  \begin{minipage}{0.5\hsize}
   \begin{center}
\hspace{-1cm}
    \includegraphics[width=6.5cm,clip]
{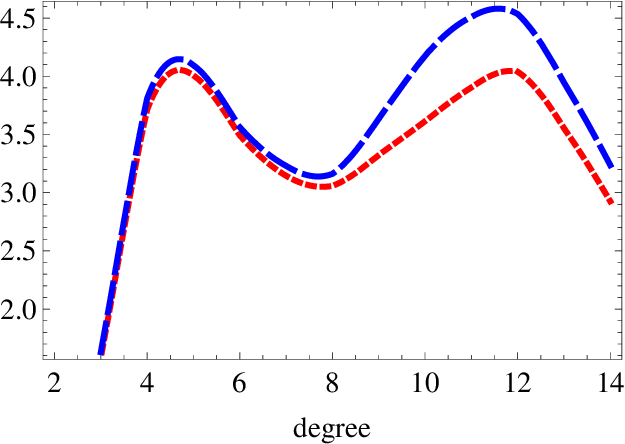}
    \end{center}
  \end{minipage}
 \end{tabular}
\vspace{0cm}
 \caption{Left: mean temperature around the center of the cold spot 
for a compensating top-hat filter (full
 curve) and for an uncompensated top-hat filter without a ring 
(dashed curve) as a function of the inner radius $\theta_{in}$. 
Right: mean temperature around the center of the cold spot divided by 
the standard deviation for the ILC map (dashed curve) and 
the Q+V+W map (full curve) as a function of $\theta_{in}$  }
\end{figure} 

As one can see in figure 12, a
deviation corresponding to the inner peak is roughly $4\, \sigma$
and that corresponding to the outer peak is $4 \sim 4.5\, \sigma$.
Assuming that the filtered mean temperature obeys Gaussian statistics, 
the statistical significances are
$P(>4\, \sigma)=6\times 10^{-3}$ and $P(>4.5\,\sigma )=
7\times 10^{-4}$ per cent.  The total solid angle of the ILC map $(|b|>20^\circ)$ is
$8.27$ sr and that of the Q+V+W map $(|b|>35^\circ)$is $5.36$ sr. 
The total number of the independent patches is roughly  
given by a ratio between the solid angle of the map and the 
area of the spherical patch with angular radius $\theta_{out}$.
Therefore, for $\theta_{out}=7^{\circ}$, we have $\sim 180$ samples
for the ILC and $\sim 110$ samples for Q+V+W map, yielding
$(110-180) \times P(>4\, \sigma)=0.7-1$ per cent. In a similar manner, 
for $\theta_{out}=17^{\circ}$, one can easily show that 
the statistical significance is $\sim 0.01-0.2$ per cent
corresponding to $\sim 3 \, \sigma$ if the likelihood function is a Gaussian
one.
 
Thus, the cold spot surrounded by a hot ring at scale $\sim 17^{\circ}$
is more peculiar than the cold spot at scale $\sim 7^{\circ}$.
Therefore, the real size of the supervoid is expected to be 
larger than the apparent size of the cold region. 
Because $1\,\sigma$ deviation (corresponding to $15-20\mu$K) 
due to a supervoid
is enough to make the signal non-Gaussian, it is reasonable to assume that
the contribution from a supervoid is less important than that from 
other effects due to acoustic oscillation or Doppler shift at
the last scattering surface.
For instance, a supervoid with a density contrast $\delta_m=-0.3$
with a comoving radius $r=200\,h^{-1}$Mpc at a redshift $z=0.2$
corresponding to an angular radius $\theta=20^\circ$
would yield a temperature decrease $\Delta T\sim 20 \mu$K in the
direction to the center. Moreover, if the supervoid is not compensating, 
a wall surrounding the supervoid could generate the observed hot ring. 
It may consist of just an ordinary underdense region surrounded by massive 
superclusters.   
Further observational study is necessary for checking the validity
of the ``local supervoid with a massive wall'' scenario.
\section{Possible solutions}
Why the observed ISW signals for prominent structures are so large? 

One possible systematic effect may come from  
a deviation from spherically symmetric density profile 
that we have not considered. Indeed, gravitational instability 
causes pan-cake or needle like structures in high density regions.
However, as we have seen, the order of the density contrast of relevant  
prominent super-structures is $\delta_m=O[0.1]$. Therefore, we expect
that the effect of anisotropic collapse plays just a minor role.
Moreover, in the case of supervoids, a deviation from spherical symmetry
is suppressed as the void expands in comoving coordinates. Thus, 
it is difficult to attribute the major cause to the deviation from spherical
symmetry. 

Another possible systematic effect is our neglect of fluctuations 
on larger scales. For instance, we may have observed just a tip of 
fluctuations whose real scale extends to $r>1000\,h^{-1}$Mpc. 
Indeed, the amplitude of the ISW effect is roughly proportional to
the scale of fluctuations, i.e.,  $\Delta T/T \propto r$ for 
$r>100\,h^{-1}$Mpc (Inoue \& Silk 2007). Therefore, the observed 
large amplitude
of ISW signal can be naturally explained. However, the angular sizes of
the observed hot and cold spots from the stacked images are just 
$4^\circ-6^\circ$ at $z\sim 0.5$ corresponding to $r=100-140\,h^{-1}$Mpc. It is difficult to explain why the 
angular sizes are so small since contributions from the ordinary 
Sachs-Wolfe effect and the early ISW effect generated near the last
scattering surface are significantly suppressed by stacking a number of 
images.

Then what are the possible mechanisms that can explain the
anomalously large ISW signals? 

One intriguing possibility
is that the primordial fluctuations are 
non-Gaussian.  Our
results suggest that the number of both supervoids and 
superclusters is significantly enhanced in comparison with the standard 
Gaussian predictions. Therefore, the effect of 
deviation from Gaussianity may appear in the statistics 
of 4-point correlations in real space or trispectrum in 
harmonic space. It can be also measured by the 
Minkowski functionals that contain information of 4-point 
or higher order correlations. At the last scattering surface,
the comoving scale of $300\,h^{-1}$Mpc corresponds to angular scale 
$\sim 2^\circ$. 
If the background universe is homogeneous, such a non-Gaussian feature 
must appear at the CMB anisotropy at multiple $l\sim 100$ corresponding
to angular scale $\sim 2^\circ$ as well. However, so far no such 
a noticeable deviation 
from Gaussianity in the CMB anisotropies 
has been observed (Vielva \& Sanz 2010). 
Therefore, it is difficult to
explain the observed signals by a simple 
non-Gaussian scenario unless one gives up the cosmological
Copernican principle (Tomita 2001). 
  
Another possibility is a certain feature on the power spectrum of primordial
fluctuation (Ichiki et al. 2009). Spike-like features
in the primordial power spectrum appear in some inflationary 
scenarios that produce primordial black holes (Ivanov et al. 1994, 
Juan et al. 1996, Yokoyama 1998). Although
there is no natural reason to have a feature only on the
scale of super-structures ($l\sim 100$), observational constraints are 
not stringent since one needs to increase the number of samples 
if one abandons an assumption of the smoothness of the primordial
power. 

Some cosmological models containing time evolving dark energy/quintessence or 
those based on scalar-tensor gravity predict an enhancement in 
the ISW effect due to an enhancement in acceleration
of the cosmological expansion or non-trivial time evolution of dark energy
or scalar field that may couple to matter or metric (Amendola 2001, Nagata et al. 2003). 
This may help to explain the
anomalously large ISW signal. However, at the same time, 
we need to suppress the
ISW contribution on large angular scales since the observed angular 
power of the CMB anisotropy 
at very large angular scales $l\sim 2$ is relatively low. 
Models based on some alternative gravity might be helpful for realizing these
observational features (Afshordi et al. 2009). 

\section{Conclusion}
In this paper, we have shown that recent observations
imply a presence of quasi-linear super-structures with a comoving 
radius $100-300\,h^{-1}$Mpc at redshifts $z<1$.  Observations are at odd with 
the concordant $\Lambda$CDM cosmology that predicts Gaussian primordial
perturbations at $>3\,\sigma$ level.

First, we have developed a formalism to estimate the amplitude of the ISW
signal for prominent structures based on thin-shell approximation
and the homogeneous collapse model. From comparison with other calculations
based on perturbation theory and the LTB solution, we 
have found that our simple model works well for estimating the 
ISW signal for quasi-linear superstructures in $\Lambda$-dominated universes.

Secondly, we have applied our developed tools to observations of the
ISW signals using the SDSS-LRG catalog, the 2MASS catalog, and 
the cold spot in the Galactic southern hemisphere in the WMAP data.
The ISW signals from stacked images for the SDSS-LRG catalog is
inconsistent with the predicted values in the concordant $\Lambda$CDM model
at more than $3\, \sigma$. The radial profiles of the stacked image
show a hot-ring around a cold spot for voids and a dip at the center of
a hot-spot without a cold-ring for clusters. These non-linear features are
also reproduced by our models using the LTB solutions although the agreement
is not perfect. The asymmetrical features suggest that the observed 
super-structures are in quasi-linear regime rather than linear regime.
The amplitudes of the ISW signals obtained from
the 2MASS catalog at redshifts $0.1<z<0.3$ are found to be several times larger
than expected values. We have confirmed that the mean temperature around
the cold spot filtered by a compensating top-hat filter with angular
scale $\theta_{out}=16^\circ-17^\circ$ deviates from
the mean value at roughly $3\,\sigma$ level suggesting a presence of a
hot-ring around the cold spot. Note that our finding is consistent 
with the previous result that the cold spot itself is not unusual
but the hot-ring plus the inner cold region is found to be
unusual (Zhang \& Huterer 2009).
This implies that a supervoid may reside
at low redshift $z<0.3$ and the angular size may be larger 
($=16^\circ-17^\circ$) than considered in literature (Masina \& Notari 2009). 

Finally, we have discussed possible causes of the discrepancy
between the theory and observation, namely, observational systematics,
primordial non-Gaussianity, features in power spectrum, dark energy or
alternative gravity. 

We have not considered effects of non-spherical collapse which are 
important for improving estimation of the mass function of non-linear objects
and effects of uncompensated mass distribution for super-structures.
The extension of our analysis to more elaborate 
ones incorporating these effects would be
helpful for realizing detailed comparison between the theory
and the observation.

Future surveys 
of the CMB, galaxy distribution, weak lensing 
and theoretical studies on dark energy/alternative gravity and
inhomogeneous cosmology will certainly yield fruitful results for 
solving the puzzle. 
\\
\\
\acknowledgements{We thank B. R. Granett for providing us
the data of radial profile of stacked images for the SDSS-LRG catalog.
We also thank I. Szapudi, H. Kodama, M. Sasaki, A. Taruya, T. Matsubara, 
and P. Cabella for useful conversation. 
Some of the results presented 
here have been derived using the Healpix
package(Go\'rski et al. 2005). 
We acknowledge the use of the Legacy Archive for Microwave Background
 Data Analysis (LAMBDA)(Lambda website).  
Support for LAMBDA is provided by the NASA Office of Space Science.
Numerical calculations were partly carried out on the 
computer center at YITP in Kyoto University.
This work is in part supported by a Grant-in-Aid for
Young Scientists (B)(20740146) and 
Grant-in-Aid for Scientific Research on Innovative Areas (22111502)
of the MEXT in Japan.} 
\appendix
\section{First-order and second-order ISW effects}
In what follows, we derive analytic formulae for computing 
temperature fluctuations due to the ISW effect for 
spherically symmetric compensated top-hat type density perturbations
using first-order and second-order perturbation theory
(Tomita \& Inoue 2008, abbreviated as TI hereafter). The relation between  
density perturbations of the growing mode and the potential function
$F$ of spatial variables are given in equations (3.6) and
(3.11) of TI. The top-hat type
density profile is parametrized in terms of two constants 
$b$ and $c$, representing first order density contrasts at the center and at
the wall at the present time (figure A1). 

The first-order density 
contrast $\delta_m^L$ at a conformal time $\eta$ when the scale factor is equal to $a(\eta)$
can be written in terms of the background matter density $\rho_B$ 
and the growth function $P(\eta)$ corresponding to the growing 
mode of density perturbation as
\BE
\delta_m^L \equiv \delta \rho^{(1)}/\rho_B = {1 \over \rho_B a^2}
[(a'/a)P' -1] (c, -b),
\EE
for $(r<r_0, r_0 <r<r_1)$, respectively, where 
a prime denotes a partial derivative with respect to conformal time
$\eta$.

The second-order density contrast is expressed as
\BE
\delta_m^S \equiv \delta \rho^{(2)}/\rho_B = {2/3\over \rho_B a^2}
\{(\zeta_1 - {9\over 2}\zeta_2) c^2, [2(c+b)^2(r_0/r)^6 +b^2] \zeta_1 +
{9\over 2}[(c+b)^2 (r_0/r)^6 -b^2]\zeta_2 \}
\EE
for $(r<r_0, r_0 <r<r_1)$, respectively, where $\zeta_1$ and $\zeta_2$
are given in equation (2.19) of TI. Here we have omitted the terms
that are negligible if $r_1 \ll H^{-1}$ because we assume that 
typical size of super-structures is $O(100)h^{-1}$Mpc.    
Neglecting the terms higher than second-order, 
the total density contrast $\delta\rho/\rho_B$ can be written as 
\BE
{\rho_B \over \rho_B +\delta \rho} = {1\over 1+ \delta \rho/\rho_B} =
1 -\delta_m^L + (\delta_m^L)^2 -\delta_m^S.
\EE


For a central value of total density contrast, $(\delta \rho/\rho)_c$, we
have the relation
\BE
\alpha (z) c^2 + \beta (z) c - \gamma (z) = 0,
\EE
where $z$ is the redshift, $\alpha (z) \equiv {2\over 3}(\zeta_1
-{9\over 2}\zeta_2)/(\rho_B a^2) - [\beta(z)]^2, \ \beta (z) 
\equiv ({a'\over a}p' -1)/(\rho_B a^2)$ \ and \ $\gamma (z) \equiv 1-
1/[1+ (\delta \rho/\rho)_c]$.

In the text we consider models of supervoids and superclusters with 
a given set of $r_0, r_1$ and $(\delta \rho/\rho)_c (z_c)$, where
$(\delta \rho/\rho)_c (z_c)$ is $(\delta \rho/\rho)_c$ at the epoch of
redshift $z_c$.
From this set we obtain $c$ and $b$, solving the above equation as 
\BE
c = - [\beta (z_c) + \sqrt{\beta (z_c)^2 + 4 \alpha (z_c) \gamma (z_c)}]/
(2 \alpha (z_c)) ,  
\EE
and $b$ is related to $c$ as $b/c = 1/[(r_1/r_0)^3 -1]$ for
compensated super-structures.

The first-order and second-order temperature fluctuations $\Delta
T^{(1)}/T$ and $\Delta T^{(2)}/T$ are defined by equations (4.2) and (4.4)
of TI. Their expressions for a light path passing the center of
spherical voids and clusters are given in equations (5.11) and (5.13) of
TI. For the other light paths, the first-order temperature fluctuation 
is derived from equations (5.8) and (5.9) with equation (C6) of TI and expressed as
\BE
\Delta T^{(1)}/T = -2 \{{a'\over a} + [{a'' \over a} - 3({a'\over
a})^2] P'\} c (r_0)^3  J(r/r_0),
\EE
where $J(r/r_0)$ is given in equation (C7) of TI for $r \le r_0$. For $r_1 >
r > r_0$, we have 
\BE
J(u) = -{1\over 6}2{u_1}^2 [I_1 + I_3 - 3{u_1}^2 I_2]/({u_1}^3 -1),
\EE 
where $u \equiv r/r_0, \ u_1 \equiv r_1/r_0$ and
\BEA
I_1 &=& \int^{u_1}_{u} (u^2 - {u}^2)^{-1/2} du = \ln [(u_1 +
\sqrt{u_1^2 - u^2})/u], \cr
I_2 &=& \int^{u_1}_{u} (u^2 - {u}^2)^{-1/2} u du =  ({u_1}^2 -
{u}^2)^{1/2}, \cr
I_3 &=& \int^{u_1}_{u} (u^2 - {u}^2)^{-1/2} u^3 du =
{1\over 3} ({u_1}^2 +2{u}^2) ({u_1}^2 -{u}^2)^{1/2}.
\EEA
The second-order
temperature fluctuation can be derived from equations (2.17),(2.18),(4.4) and
(4.5) of TI and expressed as
\BE
\Delta T^{(2)}/T = - [\zeta_1' \int^\infty_0 d\lambda (F_{,r})^2 + \zeta_2'
\times 100 \int^\infty_0 d\lambda \Phi_0 ],
\EE 
where $\int^\infty_0 d\lambda (F_{,r})^2$ and $\int^\infty_0 d\lambda
\Phi_0$ for $r<r_0$ are given in equations (C3) and (C4) of TI and the
expression of 
$\zeta_1'$ and $\zeta_2'$ is shown in equations (4.6) and (4.7) of TI.
For $r_1 > r > r_0$, we have
\BEA
\int^{\infty}_0 (F_{,r})^2 d\lambda &=&
 {1\over 9}c^2 (r_0)^3 [ {u_1}^6 I_4 + I_3 - 2{u_1}^3
I_1]/({u_1}^3 -1)^2 \cr
100 \int^{\infty}_0 \Phi_0 d\lambda &=&
{1\over 4} c^2 (r_0)^3 [{u_1}^6 I_4 -2 I_3 + 9{u_1}^2 I_2 - 8{u_1}^3
I_1]/({u_1}^3 -1)^2,
\EEA
where $I_i \ (i = 1 - 3)$ are given above and $I_4$ is
\BE
I_4 = \int^{u_1}_{u} (u^2 - {u}^2)^{-1/2} u^{-3} du =
{1\over 2{u}^2} \{{1\over u} \tan^{-1} [\sqrt{{u_1}^2
-{u}^2}/u] + \sqrt{{u_1}^2 -{u}^2}/{u_1}^2\}.
\EE

When we compare the temperature fluctuations in the perturbative model
(in appendix A) and those in the full
non-linear model (in appendix B), we should notice the difference of
their density profiles, i.e. the top-hat profile (in appendix A) and
the Sakai-Inoue (SI) profile (in appendix B). For our comparison in
this paper, we simulate the top-hat profile to the SI profile by
equating the outer boundaries and their zero points as follows. Here
we represent the SI profile using the radial coordinate $r$ defined in
the perturbative model. In the top-hat profile, the radii in the outer
boundary and the zero point are $r = r_1$ and $r_0$, respectively, and
in the SI profile the radius in the outer boundary is $r = r_{out}$ and
the zero point is $r = (r_{in} + r_c)/2$ approximately, in
which $r_c = (r_{out} + r_{in})/2$.  If we equate these outer
boundaries and zero points, we obtain   
\BE
r_0 = (r_{in} + r_c)/2 = (3 r_{in} + r_{out})/4.
\EE
Then for relative widths $w_{th} \equiv 1 - r_0/r_1$ and $w_{SI}
\equiv 1 - r_{in}/r_{out}$, we have a relation $w_{th} = {3\over 4}
w_{SI}$. In the text we show the temperature fluctuations in both
models with parameters which satisfy this relation.
\begin{figure}[t]
\vspace{-1cm}
 \begin{tabular}{cc}
  \begin{minipage}{0.5\hsize}
   \begin{center}
    \includegraphics[width=7.6cm,clip]{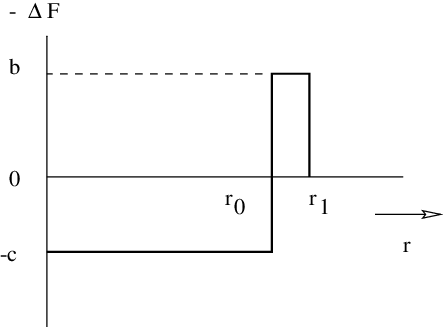}
   \end{center}
  \end{minipage}
  \begin{minipage}{0.5\hsize}
   \begin{center}
\hspace{1cm}
    \includegraphics[width=8cm,clip]{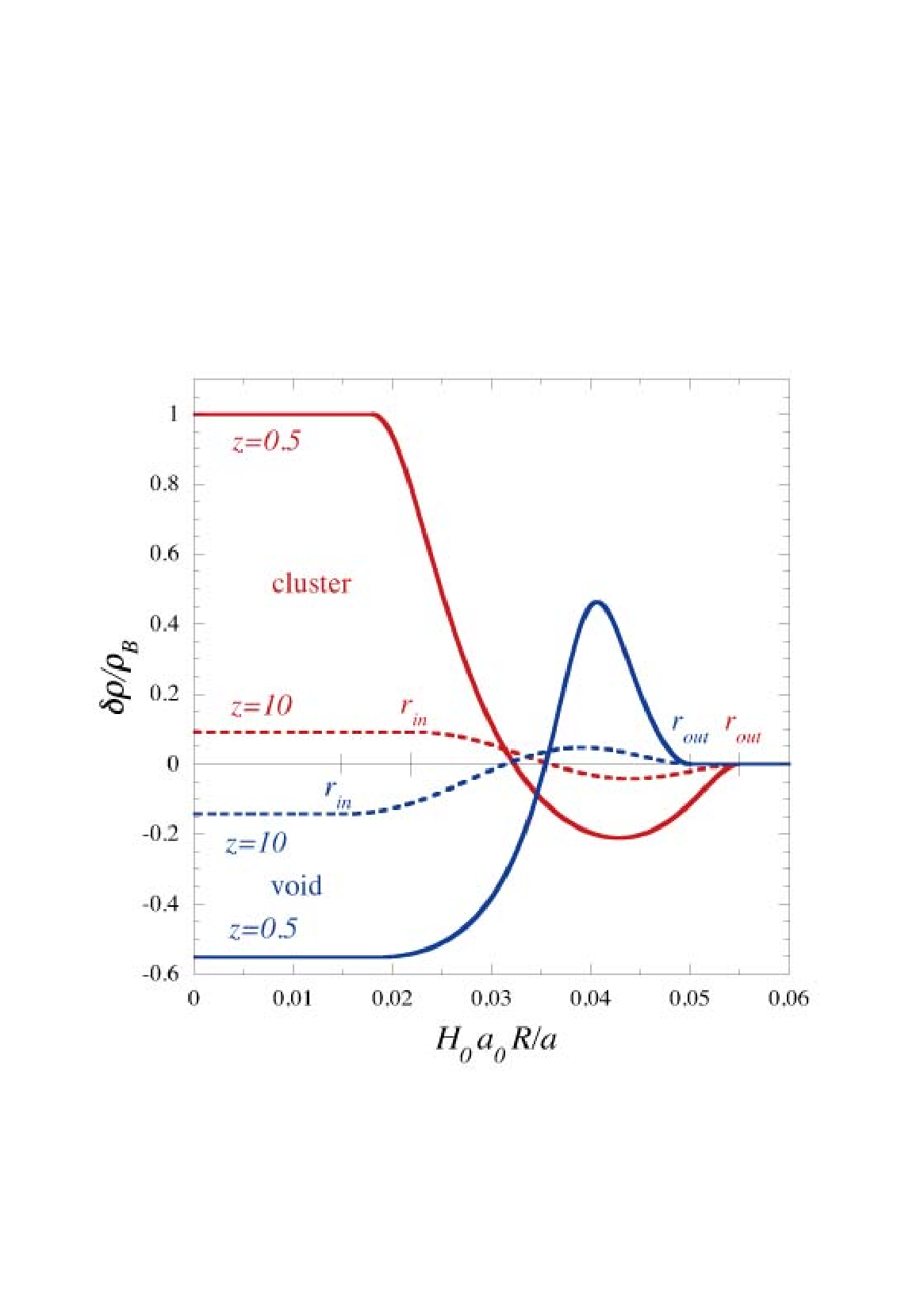}
    \end{center}
  \end{minipage}
 \end{tabular}
\vspace{-1.5cm}
 \caption{(Appendix A,B) Left: matter density 
contrast (at linear order) for a top-hat type spherical void used in 
our perturbative analysis. 
Right: density profiles $\delta_m$ in our LTB models as a function
 of physical radius in unit of present Hubble radius for 
a compensated cluster and void. The initial condition is set at a
 redshift $z=10$. We assume a concordant FRW cosmology 
with $(\Omega_{m,0}, \Omega_{\Lambda,0})=(0.26,0.74)$ as the background spacetime.} 
\end{figure} 
\vspace{1.5cm}
\section{Method of computing Rees-Sciama effects using the LTB models}

Any spherically symmetric spacetime which includes dust of energy
density $\rho(t,r)$ and a cosmological constant $\Lambda$ can be described
by the LTB solution,
\BE\label{LTBmetric}
ds^2=-dt^2+{{R'}^2(t,r)\over1+f(r)}dr^2+R^2(t,r)(d\theta^2+\sin^2\theta d\varphi^2)
\EE
which satisfies
\BE\label{LTBeq1}
\dot R^2={2Gm(r)\over R}+{\Lambda\over3}R^2+f(r)
\EE\BE\label{LTBeq2}
\rho={m'(r)\over4\pi R^2R'}
\EE
where $'\equiv\partial/\partial r$ and $ \dot{~}\equiv\partial/\partial t$.

Our model is composed of three regions: an outer flat FRW spacetime, an inner negatively/positively 
curved FRW spacetime, and an intermediate shell region described by the
LTB metric. 
At the initial time $t=t_i$, which we choose as $z_i=10$, we define the radial coordinate as $R(t_i,r)=r$,  
and we assume (figure A1)
\BE\label{m}
m(r)=\frac43\pi R^3\rho_mW(r),~~~
W(r)\equiv
\left\{\begin{array}{lcl}
1+\delta_m
&{\rm for} &r\le r_{in}\\
\displaystyle1+{\delta_m\over16}(8-15X+10X^3-3X^5)
&{\rm for} &r_{in}\le r\le r_{out}\\
1
&{\rm for} &r\ge r_{out}
\end{array}\right.
\EE
where
\BE\label{def}
X\equiv{r-r_c\over w/2},~~~
r_c\equiv{r_{out}+r_{in}\over2},~~~ 
w\equiv r_{out}-r_{in}.
\EE
Initial velocity field, $v=\dot R-HR$, is given by the linear
perturbation theory (TI).
Then $f(r)$ is determined by equation (\ref{LTBeq1}).
Our model parameters are $\Omega_{m,0}$, $r_{out}$, $w$, the redshift of the center of  a void/cluster, $z_c$, and $\delta_m(z_c)$.

The wave 4-vector $k^{\mu}$ of a photon satisfies the null
geodesic equations, 
\BE\label{geodesic}
k^{\mu}\equiv{dx^{\mu}\over d\lambda},~~~ k^{\mu}k_{\mu}=0,~~~
{dk^{\mu}\over d\lambda}+\Gamma^{\mu}_{\nu\sigma}k^{\nu}k^{\sigma}=0
\EE
For null trajectories on the $\theta=\pi/2$ plane, 
the geodesic equations (\ref{geodesic}) with the metric (\ref{LTBmetric}) reduce to
\BE\label{null}
(k^t)^2=X^2+R^2(k^{\varphi})^2,~~~
X\equiv{R'\over\sqrt{1+f}}k^r,
\EE
\BE\label{kphi}
R^2k^{\varphi}={\rm constant},
\EE\BE\label{rt}
{dr\over dt}={k^r \over k^t},~~~
{d\varphi\over dt}={k^{\varphi}\over k^t},
\EE\BE
{dk^t\over dt}=-{\dot{R'}X^2\over R'k^t}-R\dot R{(k^{\varphi})^2\over k^t},
\EE\BE\label{krt}
{dX\over dt}= -{\dot{R'}\over R'}X+R\sqrt{1+f}{(k^{\varphi})^2\over k^t}.
\EE
We use the null condition (\ref{null}) not only to set up initial data but also to check numerical precision after time-integration.

To integrate the geodesic equations (\ref{rt})-(\ref{krt}) together with the field equations (\ref{LTBeq1}) and (\ref{Yrt}) numerically, we discretize $r$ into $N$ elements,
\BE
r_i=i\Delta r, ~~~ i=1,...,N,~~~ \Delta r={r_{out}\over N}.
\EE
and any field variable $\Phi(t,r)$ into $\Phi_i(t)\equiv\Phi(t,r_i)$.
Evolution of $R_i(t)$ is determined by (\ref{LTBeq1}), but we also need data of $R'_i(t)$ and $R''_i(t)$.
Because finite difference approximation, $R'_i(t)\approx(R_{i+1}-R_{i-1})/(2\Delta r)$, include errors of $O(\Delta r^2)$, we numerically integrate with time,
\BE\label{Yrt}
\dot R'={1\over2\dot R}\left(\frac{2Gm'}{R}-\frac{2 G m}{R^2}R'+f'+\frac23\Lambda RR'\right),
\EE
which is given by differentiating (\ref{LTBeq1}) with respect to $r$.
Furthermore, to vanish $R''(t,r)$ in the geodesic equations, we have introduced an auxiliary  variable $X$.
To prepare geometrical values between grid points $r_i$ and $r_{i+1}$, 
we adopt cubic interpolation: at each time $t_*$ any variable $\Phi(t_*,r)$ in $r_i<r<r_{i+1}$ is determined by
$$
\Phi(t_*,r)=ax^3+bx^2+cx+d,~~~ x\equiv r-r_i-{\Delta r\over2}
$$$$
a\equiv{-\Phi_{i-1}+3(\Phi_i-\Phi_{i+1})+\Phi_{i+2}\over6(\Delta r)^3},~~~
b\equiv{\Phi_{i-1}-\Phi_i-\Phi_{i+1}+\Phi_{i+2}\over4(\Delta r)^2},
$$\BE
c\equiv{\Phi_{i-1}+27(-\Phi_i+\Phi_{i+1})-\Phi_{i+2}\over24\Delta r},~~~
d\equiv{-\Phi_{i-1}+9(\Phi_i+\Phi_{i+1})-\Phi_{i+2}\over16},
\EE

We also have to consider null geodesics from an observer to the void/cluster.
Suppose that the observer is at the origin and the center of the void/cluster is located at $x=x_c$ on the $y$ axis. Then, without loss of generality, on the $x$-$y$ plane we can analyze light rays which reach the observer.
Some position on the outer shell and the four momentum of the light there in the observer-centered coordinate are given by 
\BE\label{outershell}
 x=x_c+r_{out}\cos\varphi,~~~ y=r_{out}\sin\varphi,
\EE
\BE
k^\mu_{(O)}=E\left(1,-{\cos\alpha\over a},-{\sin\alpha\over a},0\right),
\EE
where $E$ is the photon energy and $\alpha$ is the angle between the light ray and the $x$-axis.
Defining $l(z)$ as a comoving length from the observer to the photon, we can write the light path as
\begin{equation}\label{lightpath}
x=l(z)\cos\alpha,\quad y=l(z)\sin\alpha,\quad
l(z)=\int^z_0{dz\over a_0H(z)}.
\end{equation}
The solution of (\ref{outershell}) and (\ref{lightpath}) gives
\BE\label{varphi}
l=x_c\cos\alpha-\sqrt{r_{out}^{~~2}-x_c^2\sin^2\alpha},\quad
\sin\varphi={l\over r_{out}}\sin\alpha,
\EE
and the null vector in the void/cluster-centered spherical coordinate,
\BE\label{k}
k^t=E,\quad
k^r=-{E\cos(\alpha-\varphi)\over r_{out}},\quad
k^{\varphi}=-{E\sin(\alpha-\varphi)\over ar_{out}}.
\EE
at the time when the photon leaves the shell, $z_{{\rm leave}}$.

Our computing algorithm is summarized as follows:
\begin{enumerate}
\item[(i)]
Suppose $\Omega_{m,0},~\delta_m(z_c),~r_{out},~w$ and redshift of the center of a void/cluster, $z_c$.
For each angle, $\alpha$, Eqs.(\ref{varphi}) and (\ref{k}) give ``initial" conditions of the null geodesic at $z=z_{{\rm leave}}$ in the void/cluster-centered spherical coordinate.
\item[(ii)]
Solve the field equations in the LTB spacetime, (\ref{LTBeq1}) and (\ref{Yrt}), from $z=z_i$ to $z_{{\rm leave}}$.
\item[(iii)]
Solve the null geodesic equations (\ref{null})-(\ref{krt}) together with the the field equations (\ref{LTBeq1}) and (\ref{Yrt}) backward from $z=z_{{\rm leave}}$ to the time when the photon enters the shell.
\end{enumerate}
\vspace{1cm}

\section{Temperature Variance for top-hat compensating filter}
In what follows, we derive analytic formulae for computing 
variance of temperature fluctuations on a sky for 
a circular top-hat compensating filter $W_{th}(\theta ;\theta_{in})$. 
We assume that an ensemble of the CMB fluctuations
can be regarded as an isotropic random field on unit sphere $S^2$.
Let $\Delta T(\theta, \phi)$ be a temperature fluctuation at 
spherical coordinates $(\theta, \phi)$. Then filterd temperature 
fluctuation centered at the ``north'' pole $(\theta=0)$ 
can be written as 
\BE
\Delta T_f=A^{-1} \int_{0}^{2 \pi}d\phi \int_0^{\theta_{out}} d \theta 
\sin{\theta}\, W_{th}(\theta;\theta_{in})\Delta T(\theta,\phi),
\label{Delta T}
\EE
where $A=2 \pi (1-\cos{\theta_{in}})$. Plugging $\Delta T$ 
expanded in spherical harmonics $Y_{lm}$,
\BE
\Delta T=\sum_{lm} a_{lm} Y_{lm}
\EE
into equation (A1), we have
\BE
\Delta T_f=A^{-1} \sum_{l} a_{l0} G_l,
\EE
where 
\BEA
G_l&=&\f{\sqrt{\pi(2l+1)}}{l+1} \biggl[2\Bigl(-x_{in}
P_l(x_{in})+P_{l-1}(x_{in})\Bigr)+x_{out}P_l(x_{out})-P_{l-1}(x_{out})
\biggr],
\nonumber
\\
x_{in}&=&\cos\theta_{in},~~ x_{out}=2x_{in}-1.
\EEA
Note that we have used a formula for the Legendre function
$P_l$, 
\BE
\f{d P_l(x)}{d x}=\f{l(l+1)}{1-x^2}\int_x^1P_l(x) dx
\EE
in deriving equation A4. Because $\Delta T$ is assumed to be isotropic on 
$S^2$, the variance of $\Delta T_f$ can be written as a function of 
the angular power spectrum $C_l$ as 
\BE
\sigma^2_f=A^{-2}\sum_lC_l G_l^2.
\EE

If the CMB sky is smoothed by a Gaussian beam with the FWHM $\theta_s$,
then the variance is 
\BE
\sigma^2_f=A^{-2}\sum_lC_l B_l G_l^2,
\EE
where $B_l=\exp[{-\sigma_s^2l(l+1)/2}]$, and $\sigma_s=(8 \ln{2})^{-1/2}\theta_s$.

In the absence of complete sky coverage, we cannot directly observe
$C_l$.  We can only compute estimated expansion coefficients for 
the observed region $R$ in the sky (Bunn 1995), 
\BE
b_{lm}=N_{lm}\int_R \Delta T Y_{lm} \sin{\theta}\, d\theta d\phi,
\EE 
where $N_{lm}$ is a factor chosen to normalize $b_{lm}$ appropriately.
If $R$ is azimuthally symmetric, one possible prescription is to set
(Peebles, 1973)
\BE
N_{lm}=W_{llm}^{-1/2},
\EE
where
\BE
W_{l l' m}=\int_R Y_{lm}Y_{l'm}\sin{\theta}\, d\theta d\phi.
\EE
Then a possible estimator for $C_l$ is given by
\BE
\tilde{C}_l \equiv\langle |b_{lm}|^2 \rangle=\f{1}{2l+1}\sum_{l',m}W_{llm}^{-2}
W_{ll'm}^2 C_{l'}.
\EE
In the limit that $C_l$ varies much more slowly than $W_{ll'm}$
\BE
\f{1}{C_l} \f{d C_l}{dl}\ll \f{1}{W_{l l' m}}\f{d
W_{ll'm}}{dl'}\bigg{|}_{l=l'},
\EE
we have $\tilde{C}_l\approx C_l$.

\end{document}